%% file: Reachability_fmcad19.tex
\documentclass[conference]{IEEEtran}
\usepackage{times}
\usepackage{soul}
\usepackage{url}
\usepackage[utf8]{inputenc}
\usepackage[small]{caption}
\usepackage{graphicx}
\usepackage{amsmath}
\usepackage{booktabs}
\usepackage{algorithm}
\usepackage[noend]{algcompatible}
\usepackage{dsfont}
\usepackage[capitalize,noabbrev]{cleveref}
\usepackage{comment}
\usepackage[detect-all,per-mode=symbol]{siunitx}
\DeclareSIUnit\feet{ft}
\usepackage[keeplastbox]{flushend}

\newcommand{\citet}[1]{\citeauthor{#1}~\cite{#1}}

\newif\ifmarkchanges
\newcommand{\changed}[1]{
	\ifmarkchanges
	\textcolor{blue}{#1}
	\else
	#1
	\fi
}

\usepackage{pgfplots}
\pgfplotsset{compat=newest}
\pgfplotsset{every axis legend/.append style={%
		cells={anchor=west}}
}
\pgfplotsset{every y tick label/.append style={font=\footnotesize}}
\pgfplotsset{every x tick label/.append style={font=\footnotesize}}
\pgfplotsset{every axis x label/.append style={font=\footnotesize}}
\pgfplotsset{every axis y label/.append style={font=\footnotesize}}
\pgfplotsset{every axis legend/.append style={font=\footnotesize}}
\pgfplotsset{every axis title/.append style={font=\footnotesize}}
\usepgfplotslibrary{polar}
\usetikzlibrary{arrows}
\tikzset{>=stealth'}
\usepgfplotslibrary{groupplots}
\usepackage{aircraftshapes}

\usepackage[backend=bibtex,style=ieee]{biblatex}
\begin{document}
\markchangesfalse
	
\title{A Reachability Method for Verifying Dynamical \\ Systems with Deep Neural Network Controllers}

\author{
	Kyle D. Julian \And
	Mykel J. Kochenderfer
	\affiliations
	Department of Aeronautics and Astronautics, Stanford University, Stanford, CA 94305
	\emails
	\{kjulian3, mykel\}@stanford.edu
}

\author{\IEEEauthorblockN{Kyle D. Julian}
	\IEEEauthorblockA{Department of Aeronautics and Astronautics\\
		Stanford University\\
		Stanford, CA 94305\\
		Email: kjulian3@stanford.edu}
	\and
	\IEEEauthorblockN{Mykel J. Kochenderfer}
	\IEEEauthorblockA{Department of Aeronautics and Astronautics\\
		Stanford University\\
		Stanford, CA 94305\\
		Email: mykel@stanford.edu}}

\maketitle

\begin{abstract}
  Deep neural networks can be trained to be efficient and effective controllers for dynamical systems; however, the mechanics of deep neural networks are complex and difficult to guarantee. This work presents a general approach for providing guarantees for deep neural network controllers over multiple time steps using a combination of reachability methods and open source neural network verification tools. By bounding the system dynamics and neural network outputs, the set of reachable states can be over-approximated to provide a guarantee that the system will never reach states outside the set. The method is demonstrated on the mountain car problem as well as an aircraft collision avoidance problem. Results show that this approach can provide neural network guarantees given a bounded dynamic model.
\end{abstract}

\section{Introduction}
Neural networks are global function approximators that can efficiently and accurately represent any target domain. In recent years, neural networks have become widely used for decision making and control in dynamical systems such as pendulums~\cite{riedmiller2005neural}, video games~\cite{mnih2015human}, autonomous vehicles~\cite{bojarski2016end}, and aircraft collision avoidance systems~\cite{julian2016policy}. Neural network controllers can be generated in many ways, including supervised learning to compress existing controllers~\cite{julian2017neural}, Neural Fitted Q Iteration~\cite{riedmiller2005neural}, and deep Q-learning~\cite{mnih2015human}.

In some domains, neural network controllers can make decisions better than human experts, but guarantees about their performance are difficult to prove, inhibiting their use in safety-critical systems. Whereas traditional model checking techniques may be sufficient to verify logic-based controllers~\cite{mcmillan1993symbolic}, new verification techniques are required to ensure that complex and non-linear neural network controllers will always generate safe trajectories. 

Recent advancements in neural network verification tools have yielded many options for proving bounds on neural network outputs~\cite{liu2019algorithms}. Given some region of the input space, these verification tools define what values could be given by the neural network and provide either a guarantee of a correct output or a counterexample. Some tools work layer by layer through the network to compute a reachable set of each layer, ending with the reachable output of the network given the input region~\cite{gehr2018ai2,xiang2017reachable}. Other tools formulate neural network verification as a mixed-integer linear programming and use optimization methods to find counterexamples~\cite{lomuscio2017approach,tjeng2017evaluating}. Tools such as Reluplex~\cite{katz2017reluplex} and Sherlock~\cite{dutta2018output} convert neural networks into linear programs and systematically search for a satisfying counterexample. While these tools ensure that input regions satisfy some network output property for a single evaluation, ensuring safety for multiple network evaluations over time remains difficult due to system dynamics and uncertainties. 

Recent work involved developing methods for the verification of neural network controllers for agents acting in an environment. These methods use a reachability approach that computes the set of states the system can reach given an initial set of states and a controller. \citet{akintunde2018reachability} formulate neural agents and their dynamics as a large mixed-integer linear program, which can be solved to verify the agent's safety. This method requires linearly-definable dynamics, but non-linear dynamics can be approximated by a neural network with arbitrary precision. \citet{ivanov2018verisig} formulate sigmoid-activated neural networks with one hidden layer as a hybrid system to compute reachability sets over time. 
\citet{xiang2019reachable} train neural networks to output the next state in a nonlinear system instead of a control input. They compute an over-approximation of the reachable set at each time step to bound the trajectories of state variables. 
\changed{In addition, \citet{xiang2018reachability} bound the neural network output over small hyper-rectangular input sets and use reachability tools for linear systems to compute reachable sets for a trajectory.}

This work builds upon prior work and \changed{uses a general} reachability algorithm for an agent controlled by a neural network. Whereas \changed{many} previous approaches combine the neural network controller output with a dynamics model, we separate the two components and analyzes each individually. Our method is applicable to any network architecture or activation function, and any existing neural network verification tool can be used. Furthermore, by analyzing the dynamics separately, our method can consider complex actions that are not well represented by linear models. \changed{While \citet{xiang2018reachability} demonstrate their approach on continuous-time linear control systems, our experiments use discrete control actions. Additionally, our} work considers the degree of uncertainty that can be present in the system dynamics before safety can no longer be guaranteed, allowing us to make claims about how a real system with errors will be able to maintain safety. A mountain car example is provided to illustrate the approach on a classic problem. In addition, an aircraft collision avoidance problem \changed{inspired by the next-generation system ACAS X~\cite{Kochenderfer2015chapter10}} demonstrates how the approach can handle complex \changed{and real-world} systems. \changed{Our experimental results provide performance and safety guarantees from any initial starting point in the full input space.}

\section{Approach}
This section describes the reachability method used to provide guarantees about neural network controlled systems. First, the neural network is over-approximated in order to bound the outputs given in different regions of the state space. Then a general reachability method is formulated that uses the network bounds to over-approximate the system's set of reachable states, which provides a guarantee that states outside the set cannot be achieved. If unsafe states lie entirely outside the reachable set, then the system is safe.

\subsection{Neural Network Over-approximation}\label{sec:netApprox}
Neural network controllers give control actions $a=f(s)$ for states $s \in \mathcal{S}$, where $\mathcal{S}$ is the controller state space. Because the neural network function can be highly nonlinear, predicting the neural network output can be difficult. However, many tools have been developed recently to verify neural networks, which can bound the behavior of neural networks. For a small set of states $S \subset S$,
tools such as Reluplex~\cite{katz2017reluplex}, Sherlock~\cite{dutta2018output}, Reluval~\cite{wang2018formal}, and Verisig~\cite{ivanov2018verisig} can be used to guarantee that $f(s) \in \mathcal{A}_S \; \forall \; s \in S$, where $\mathcal{A}_S$ is a bound on the neural network output in set $S$. For neural networks with discrete outputs, $\mathcal{A}_S$ represents the set of all discrete actions that could be chosen by the neural network for some point in $S$. For neural networks with $n$ continuous control outputs, $\mathcal{A}_S$ represents some $n$-dimensional polytope. If $\mathcal{S}$ is divided into $N$ small subsets, or cells, then $\mathcal{A} = \{\mathcal{A}_{S_1},\mathcal{A}_{S_2},\dots,\mathcal{A}_{S_N}\}$ bounds the neural network output. By assuming that the neural network can give any action $a \in \mathcal{A}_S$ from any point $s \in S$, the neural network controller is over-approximated. 

\subsection{Reachability Method}
Let $\mathcal{S}_N = \{S_1,S_2,\dots,S_N\}$ be the set of all $N$ cells that divide $\mathcal{S}$, and define $\mathcal{R}_t \subseteq \mathcal{S}_N$ to be the set of cells reachable at time $t$. The initial reachable set $\mathcal{R}_0$ is problem specific and can include all or some specific subset of $\mathcal{S}_N$. The reachability method iteratively computes $\mathcal{R}_{t+1}$ given $\mathcal{R}_{t}$ and $\mathcal{A}$. To compute $\mathcal{R}_{t+1}$, first compute the region of state space $R_S \subseteq \mathcal{S}$ that can be reached at the next time step from $S$ with control inputs from $\mathcal{A}_S$. With system dynamics $g(s,a)$ for state $s$ and control input $a$, $R_S$ can be defined as 

\begin{equation}
R_S = \{g(s,a) : s \in S, a \in \mathcal{A}_S \}
\end{equation}

In general, there may be uncertainty surrounding $g(s,a)$. Factors such as modeling errors, disturbances, or human error may cause the dynamics of real systems to differ from ideal equations. This work considers these errors to obtain safety guarantees considering some level of error in $g(s,a)$. One way to incorporate these errors is to expand $R_S$ to include all states that could be reached with error in the dynamics.

Next, the set of cells that could be reached from cell $S_i \in \mathcal{R}_{t}$ is calculated as 
\begin{equation} \label{eq:reachSet}
\mathcal{R}^{s_i}_{t+1} = \{ S : S \in \mathcal{S}_N , S \cap R_{S_i} \neq \emptyset \}
\end{equation}
\Cref{eq:reachSet} over-approximates $R_S$ by including all cells that have some intersection with $R_S$. Lastly, the set of cells reachable at the next time step from any cell in $\mathcal{R}_t$ is 
\begin{equation}
\mathcal{R}_{t+1} = \bigcup_{S_i \in \mathcal{R}_{t}} \mathcal{R}^{S_i}_{t+1}
\end{equation}
\cref{alg:algorithm} outlines the steps for this method.

\begin{algorithm}[tb]
	\caption{Reachability for networks with discrete actions}
	\label{alg:algorithm}
	\textbf{Input}: $\mathcal{R}_0$, $\mathcal{S}_N$, $T$\\
	\textbf{Output}: $\{\mathcal{R}_{1},\mathcal{R}_{2},\dots,\mathcal{R}_{T}\}$
	\begin{algorithmic}[1] %[1] enables line numbers
		\FOR{$t=1$ to $T$}
		\STATE $\mathcal{R}_{t} \leftarrow \emptyset$
		\FOR{$S \in \mathcal{R}_{t-1}$}
		\STATE Compute $\mathcal{A}_S$ using neural network verification tool
		\FOR{$a \in \mathcal{A}_S$}
		\STATE Compute $R_{S}$ using state dynamics
		\FOR{$S' \in \mathcal{S}_N$}
		\IF{$S' \bigcap R_{S} \neq \emptyset$}
		\STATE $\mathcal{R}_t \leftarrow \mathcal{R}_t \bigcup S'$
		\ENDIF
		\ENDFOR
		\ENDFOR
		\ENDFOR
		\ENDFOR
		\STATE \textbf{return} $\{\mathcal{R}_{1},\mathcal{R}_{2},\dots,\mathcal{R}_{T}\}$
	\end{algorithmic}
\end{algorithm}

Because the neural network and system dynamics are over-approximated, this method will over-approximate the reachable set at each time step. As a result, all states outside the reachable set at time $t$ are impossible to reach from the initial set of states. If the reachable set is contained within a goal set or excludes a failure set, then this method verifies that the neural network controller will be safe at time $t$.  

\subsection{Implementation Considerations}
Although the proposed method can verify neural network controllers, its implementation may raise a few concerns. First, neural network verification tools can be slow, so checking each discrete cell may seem slow especially as the number of cells grows. However, many neural network verification tools run faster as input bounds become tighter, so checking many small regions may not be slower than checking fewer large regions. Furthermore, computing $R_S$ becomes more complicated with higher-dimensional state spaces. In two dimensions with linear dynamics and rectangular cells, computing $R_S$ is straightforward, but more complicated dynamic models for higher-dimensional spaces can prove challenging. Fortunately many tools exist to perform reachability over-approximations for complex problems, such as SpaceEx, which uses lazy set computation with Zonotopes for efficient computation with hundreds of state variables~\cite{frehse2011spaceex}. Another tool, JuliaReach, uses the high performance Julia language to compute reachable sets assuming convex sets and currently supports linear dynamics~\cite{bogomolov2019juliareach}. These tools provide efficient methods for reachable set estimation that can be used to enable this approach for complex neural network systems.

Additionally, if a cell $S \in \mathcal{R}_t$ is also in $\mathcal{R}_{t+1}$ given control inputs $\mathcal{A}_S$, then $S$ will remain in all future reachable sets, which could be problematic if $S$ is in an undesirable area of the state space. This problem can be caused by over-approximation errors, which can be improved by shrinking the cells. However, making cells smaller can quickly lead to an explosion in the number of cells to consider, especially at higher dimensions. It is important to reduce cell size in critical areas of the state space without generating so many cells that reachability computation becomes too slow.

\section{Mountain Car Example}

\begin{figure}
	\centering
	\includegraphics{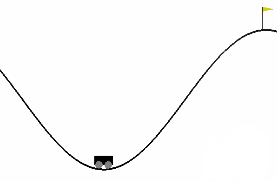}
	\caption{\changed{Mountain car problem~\cite{openai} with goal state marked by the flag}}\label{fig:mc}
\end{figure}

To demonstrate the approach, we consider the classic mountain car problem depicted in \cref{fig:mc}, where a one-dimensional car must climb to the top of a hill~\cite{moore1990efficient}. However, the car's acceleration cannot overcome gravity, so the car must rock between two hills to gain enough momentum to reach the top of the hill. The system has two state variables, position $p$ and velocity $v$, and one control input, $u$. We consider the same dynamic model as used by \citet{ivanov2018verisig} but with \changed{uniformly distributed random} disturbance $\delta_u \sim \mathcal{U}(-w,w)$ added to the control input, where $w$ is the maximum disturbance. The state variables are updated as
\begin{equation}
\begin{bmatrix}
p \\
v
\end{bmatrix}
\leftarrow
\begin{bmatrix}
p+v \\
v + 0.0015(u+\delta_u) - 0.0025 \cos(3p) 
\end{bmatrix}
\end{equation}

This formulation uses discrete actions, $u \in \{-1,0,1\}$, instead of continuous values. The range of positions and velocities is $p \in [-1.2,0.6]$ and $v \in [-0.07,0.07]$ respectively, and the goal of the car is to reach $p=0.6$ as quickly as possible. 

\subsection{Neural Network Controller}\label{sec:nnet}

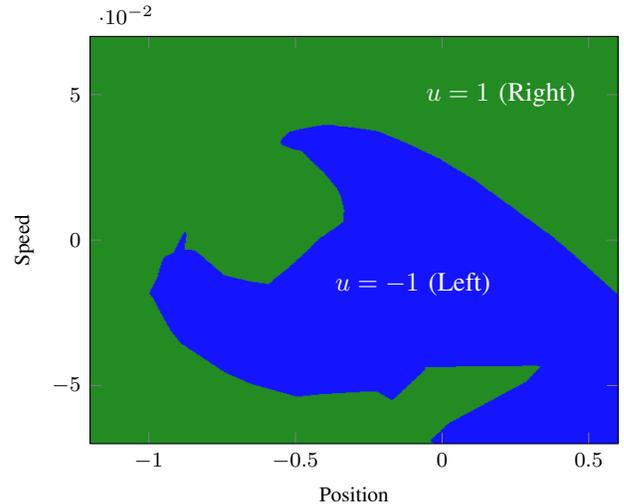
\begin{figure}
	\centering
	\input{MountainCarPolicy.tex}
	\caption{Mountain car neural network controller policy}
	\label{fig:MCpolicy}
\end{figure}

This section presents a method for generating neural network controllers, though there are many other methods that could be used with this reachability approach. To generate a neural network controller, the problem was first solved as a discrete Markov decision process using POMDPs.jl~\cite{egorov2017pomdps}. In a Markov decision process, an agent in state $s \in \mathcal{S}$ takes action $a \in \mathcal{A}$, receives reward $r(s,a)$, and transitions to a new state $s'$ with probability $T(s,a,s')$, where $T$ is the transition function~\cite{Kochenderfer2015chapter4}. Discrete value iteration was used to compute policy $\pi$ that maximizes the accumulation of reward overtime. In discrete value iteration, $Q$-values are associated with each state-action pair, with initial $Q_0(s,a)=0$. The $Q$-values are updated according to the finite-horizon Bellman equation as 
\begin{equation}
Q_{t+1}(s,a) = r(s,a) + \max_{a'} \sum_{s'}T(s,a,s')Q_t(s',a')
\end{equation}
After the $Q$-values converge, the policy is computed as $\pi(s) = \arg\max_{a} Q(s,a)$. For the mountain car problem, $s=(p,v)$, $\mathcal{S}$ is the Cartesian product of 100 uniformly distributed positions and velocities, $a = u$, $r(s,a)=-\mathds{1}(p<0.6)$, and $T(s',a,s)$ has $1/3$ probability for each of the next states calculated with $\delta_u$ as -0.5, 0.0, and 0.5, which encourages the computed policy to be robust to noisy accelerations. Multilinear interpolation is used to compute the $Q(s',a')$ when $s'$ does not fall on one of the points in the grid.

After computing the $Q$-values for the optimal policy, a neural network is trained through supervised learning to approximate the $Q$-values. The neural network has an input variable for each state dimension and an output variable associated with each action. An asymmetric loss function that penalizes under-valuing the optimal action or over-valuing the suboptimal actions was used, which encourages the network to approximate both the $Q$-values and policy well~\cite{julian2016policy}. For the mountain car problem, the neural network uses ReLU activations and has five hidden layers of 30 units each. The neural network was trained for 1000 epochs using Adamax optimization~\cite{Adam}, and the trained network predicts actions with 97.80\% accuracy and an average absolute error of 5.69. The the policy of the trained network is shown in \cref{fig:MCpolicy}, which shows that the neural network does not recommend $u=0$ actions often if at all.

\subsection{\changed{State Space Discretization}}
\begin{figure}
	\centering
	\input{GridDensity.tex}
	\caption{\changed{Grid discretization for mountain car problem}}
	\label{fig:MC_density}
\end{figure}
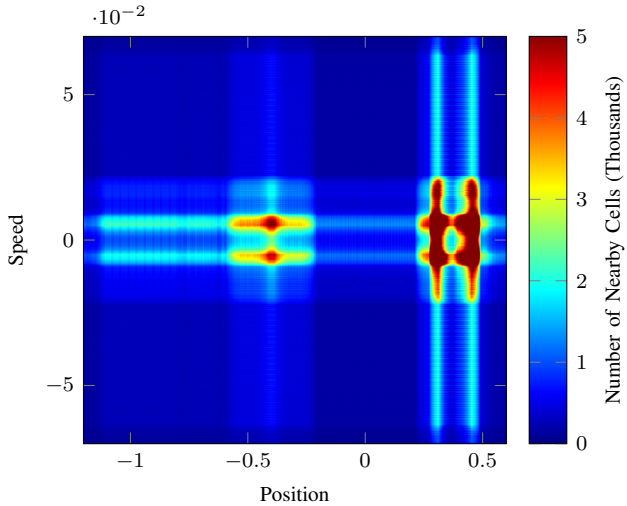
\changed{
The discretization of the input space into cells has a significant impact on the performance of this algorithm. First, a larger number smaller cells requires checking more queries to with neural network verification tools, which may increase computation time to approximate the neural network controller. In addition, using larger numbers of cells will increase the amount of computation required to compute reachable sets. However, larger cells lead to more over-approximation errors, which can prevent the algorithm from proving the desired closed-loop property. For the mountain car problem, we wish to show that all initial starting conditions will reach the goal. If over-approximation errors result in a cell being in its reachable set at the next time step, then we cannot prove that the controller will move the car away from the cell and towards the goal. Therefore, if a cell is within its reachable set, the cell should be divided until it is no longer a member of its reachable set. 
}

\changed{
Selecting a good state space discretization is important for many reachability algorithms. One method for discretization uses a tree representation and refines leaf nodes as needed~\cite{han2008approximate}. Similarly, we iteratively refined the grid when cells were computed to be in their own reachable set. \Cref{fig:MC_density} shows a heat map of the final discretization used for the mountain car problem, where the value at each pixel corresponds to the number of grid cells within a small region around the pixel. The discretization is uniform in most locations, but a few key areas are more heavily discretized. First, regions with low magnitude velocity are more heavily discretized because the position does not change much when velocity is small. In addition, the region around $p=-0.45$ is refined because this location is the bottom of the hill where the $\cos(3p)$ gravity term is small. Lastly, the region around $p=0.4$ is refined. In this region, selecting the action to move right has just enough power to overcome the gravity term, resulting in a slow climb up the hill. However, if this region is not finely discretized, then over-approximation errors would show the car getting stuck on the side of the hill if the right action is chosen. The final state space discretization contains 341523 hyper-rectangular cells.}

\subsection{Neural Network Approximation}

\begin{table}
	\centering
	\caption{Mountain car neural network approximation results}
	\begin{tabular}{lrrr}  
		\toprule
		Method  & Num. Actions & Runtime (s) & Complete \\
		\midrule
		Sampling     &  344126  & 172.9   & No  \\
		ReluVal      &  344127  & 72.99   & Yes  \\
		Reluplex     &  344130  & 57360   & Yes \\
		
		\bottomrule
	\end{tabular}
	\label{tab:MCtable}
\end{table}

\begin{figure*}
	\centering
	\input{MC_ReachableRow.tex}
	\caption{Mountain car reachable set over time without disturbances}
	\label{fig:MC_reachRow}
\end{figure*}
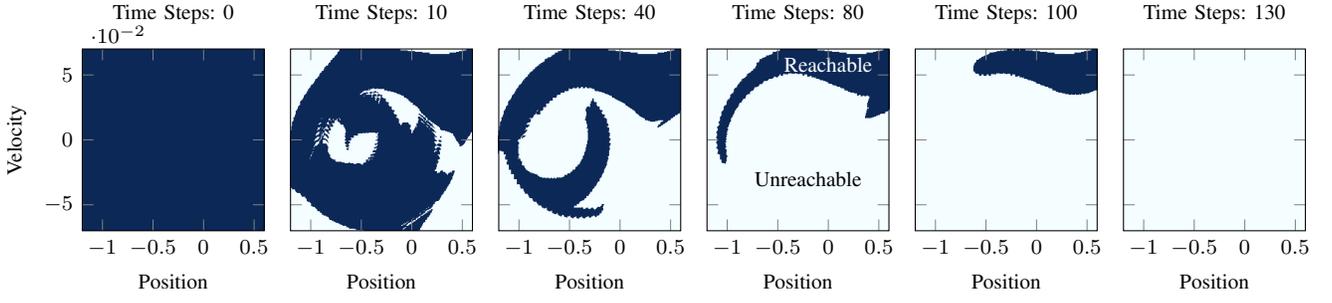

Three methods were used to approximate the neural network policy as described in \cref{sec:netApprox}: 100 random samples, Reluplex~\cite{katz2017reluplex}, and ReluVal~\cite{wang2018formal}. 
\changed{Reluplex is an SMT solver that extends the simplex method to handle ReLU activations and check if a set of input variables in the input region can satisfy an output constraint~\cite{katz2017reluplex}.}
\changed{ReluVal uses symbolic interval analysis and iteratively tightens the bound until the output property is verified or a counterexample is found~\cite{wang2018formal}.}
\changed{To use Reluplex and ReluVal, the property we want to hold must be negated and checked for a counterexample. For a discrete action neural network, the action associated with the highest network output value is used. Therefore, to check if an action is given, the output constraint is that the value of the target action must be less than the value of all other actions. If a counterexample is found, then there exists a point in the input set where the target action is given. Otherwise, if the solvers return UNSAT, then we are guaranteed the target action is not given within the input region. For every input region to be checked, we target each possible action to see if that action could be given by the neural network.}

\Cref{tab:MCtable} reports the results where the number of actions is the sum of the number of actions found in all cells, and the run time reported uses a single CPU, though all methods are easily parallelized to decrease runtime. The sampling method misses an action in one of the cells, and Reluplex finds three spurious actions that occur very close to the cell but outside the cell boundaries. Decreasing the boundary error tolerance removes these spurious actions. ReluVal identifies all actions given within each cell and performs faster than both Reluplex and sampling approaches.

\begin{figure}
	\centering
	\input{ReachableCells.tex}
	\caption{\changed{Mountain car reachable sets and randomly sampled points from initial cell}}
	\label{fig:reach}
\end{figure}
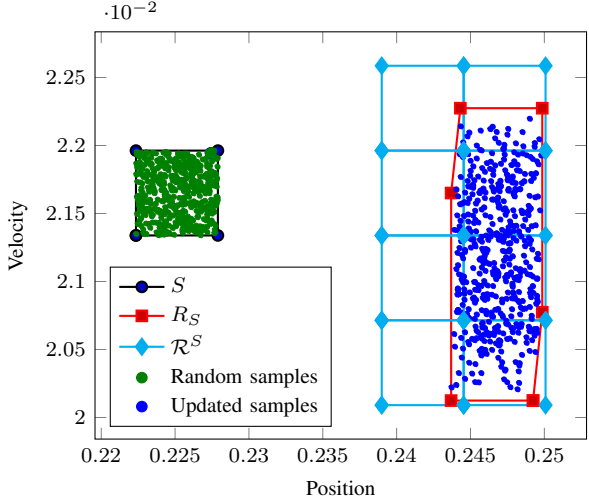

\subsection{Reachability Results}
\changed{
First, we need to compute bounds on the reachable set of next states $R_S$ from a given set of initial states $S$. For a given hyper-rectangular initial set, the input variables are bounded by $p \in [p_{\text{min}},p_\text{max}]$ and $v \in [v_\text{min},v_\text{max}]$, and the disturbance term is bounded by $\delta_u \in [-w,w]$. Although the dynamics include a nonlinear $\cos(3p)$ term, we can bound the function as 
\begin{equation}
\cos(3p) \in \left[cp_\text{min},cp_\text{max}\right]
\end{equation}
where
\begin{align}
cp_\text{min} &= \min_{p\in[p_\text{min},p_\text{max}]}\cos(3p) \\
cp_\text{max} &= \max_{p\in[p_\text{min},p_\text{max}]}\cos(3p)
\end{align}
}

\changed{
As a result, we can compute minimum and maximum bounds for the state variables at the next time step $(p',v')$ as 
\begin{align}
p' &\ge p_\text{min}+v_\text{min} \\
p' &\le p_\text{max} + v_\text{max} \\
v' &\ge v_\text{min} - 0.0025 cp_\text{max} + 0.0015 (u-w) \\
v' &\le v_\text{max} - 0.0025 cp_\text{min} + 0.0015(u+w)
\end{align}
Lastly, by subtracting $p'$ from $v'$, we get $v'-p'=-p -0.0025\cos(3p)+0.0015(u+\delta_u)$. We can bound this term as 
\begin{align}
v'-p' &\ge -p_\text{max} - 0.0025 cp_\text{max} + 0.0015 (u-w) \\
v'-p' &\le -p_\text{min} - 0.0025 cp_\text{min} + 0.0015 (u+w)
\end{align}
}

\changed{
Using these six inequalities, we can compute a convex polygon that contains all possible variables at the next time step from some initial set. \Cref{fig:reach} shows an example with $w=0.5$ and $u=1.0$. In addition, 500 random points are initially sampled from the input region and updated with the mountain car dynamics. All 500 points are bounded by the convex polygon $R_S$. Then, $\mathcal{R}^S$ is derived as the set of cells in the state space that overlap with $R_S$.
} 

Initially, the car could be in any state, so $\mathcal{R}_0 = \mathcal{S}_N$. \changed{For each cell in $\mathcal{R}_0$, the set of cells that could be reached at the next time step is computed as illustrated in \Cref{fig:reach}, and $\mathcal{R}_1$ is computed as the union of reachable cells for any of the cells' possible actions. \Cref{fig:MC_reachRow} shows the evolution of over-approximated reachable set over time using the neural network controller.}  Over time the reachable set diminishes to include only states where $p=0.6$, which proves that the car will always reach the goal when using the neural network controller regardless of initial condition. \changed{The runtime required to evolve reachable sets is on the order of a couple seconds, much faster than the time required to over-approximate the neural network controller.}

\begin{figure}
	\centering
	\input{MC_reach.tex}
	\caption{Mountain car maximum time to reach goal with disturbances}
	\label{fig:MC_reach}
\end{figure}
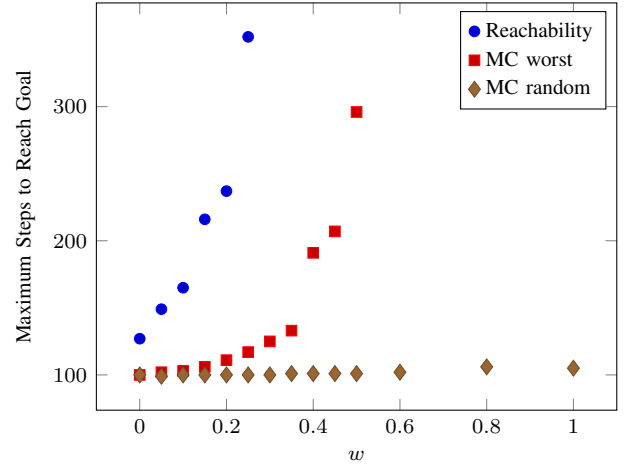

Reachable sets were also computed with different levels of dynamic disturbance to study how errors in the dynamics effect verification. The reachability method was compared to two Monte Carlo simulations that use worst-case and random disturbances. The worst-case disturbance is $\delta_u=-\text{sign}(u)w$ so that the disturbance always opposes the car's acceleration, while random disturbances are sampled from $\mathcal{U}(-w,w)$ at each time step. For both Monte Carlo methods, 10000 simulations from random starting locations were evaluated, and the maximum number of time steps to reach the goal state was recorded. \Cref{fig:MC_reach} summarizes the results and shows that the reachability method guarantees that the car will reach the goal up to disturbance level $w=0.25$; however, worst-case Monte Carlo simulations show that the car will reach the goal for $w\le0.5$. Over-approximation errors contribute to the size of the gap between the two approaches, and decreasing the cell sizes to reduce over-approximation error could decrease this gap. Random Monte Carlo simulations show no difference for different levels of disturbances. Since the disturbance distribution has zero mean, the disturbance has no effect in expectation. Verifying safety in critical systems means ensuring failures do not occur even with low probability, which requires firm bounds on dynamic uncertainties to prevent large disturbances from having non-zero probability.

\section{Aircraft Collision Avoidance System Example}

The second example considers commercial aircraft, which are required to operate with a collision avoidance system that gives vertical climbrate advisories to pilots to prevent near midair collisions (NMACs). An NMAC occurs when the aircraft are separated by less than \SI{100}{\feet} vertically and \SI{500}{\feet} horizontally. 
An aircraft collision avoidance system called ACAS X uses $Q$-values to represent much of the decision making logic~\cite{Kochenderfer2015chapter10}. It uses a large table of state-action $Q$-values to make decisions, but \citet{julian2016policy} explored compressing the table using a neural network. This section presents a notional example based loosely on an early prototype of ACAS X~\cite{Kochenderfer2015chapter10}.

\subsection{System Description}
The example collision avoidance system, referred to as VerticalCAS, considers two aircraft: an ownship \changed{aircraft} equipped with VerticalCAS, and an intruder aircraft. In this formulation, the intruder is assumed to maintain level flight, but future work can relax this assumption. The system uses four variables to describe the encounter with the intruder aircraft:
\begin{enumerate}
	\item $h$ (\si{\feet}): Intruder altitude relative to ownship $[-3000, 3000]$
	\item $\dot{h}_0$ (\si{\feet\per\minute}): Ownship vertical climbrate $[-2500, 2500]$
	\item $\tau$ (\si{\second}): Time to loss of horizontal separation $[0, 40]$
	\item $s_{\text{adv}}$: Previous advisory from VerticalCAS
\end{enumerate}

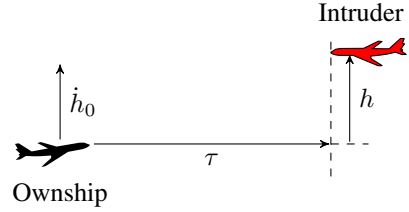
\begin{figure}
	\centering
	\begin{tikzpicture}
	\node [aircraft side,fill=black,minimum width=1cm,rotate=10] (own) at (0,0) {};
	\coordinate[label=below:Ownship] (ownText) at (0,-0.3);
	\node [aircraft side,draw=black,fill=red,minimum width=1cm,xscale=-1] (int) at (4,1.3) {}; 
	\coordinate[label=above:Intruder] (intText) at (4,1.6);
	
	\draw (0.45,0.09) edge[->] node[midway,below] {$\tau$} (3.58,0.09);
	\draw [dashed] (3.6,1.45) -- (3.6,-0.35);
	\draw [dashed] (3.6,0.09) -- (4.1,0.09);
	\draw (0.0,0.15) edge[->] node[midway,right] {$\dot{h}_0$} ++(0.0,1.0) ;
	\draw (3.85,0.12) edge[->] node[midway,right] {$h$} ++(0.0,1.16) ;
	\end{tikzpicture}
	\caption{\changed{Aircraft encounter geometry for VerticalCAS example}}\label{fig:VerticalCAS}
\end{figure}

The first two state variables describe the encounter geometry vertically. The $\tau$ variable condenses the horizontal geometry into a single variable by providing a countdown until the intruder will no longer be separated horizontally, at which point the ownship must be vertically separated to avoid an NMAC. These geometric variables are depicted in \cref{fig:VerticalCAS}.

The $s_{\text{adv}}$ variable is categorical and can be any one of the nine possible advisories given by the system, and conditioning the next advisory on the current advisory allows the system to maintain consistency when alerting pilots. The nine possible advisories are 
\begin{enumerate}
	\item COC: Clear of Conflict
	\item DNC: Do Not Climb
	\item DND: Do Not Descent
	\item DES1500: Descend at least \SI{1500}{\feet\per\minute}
	\item CL1500: Climb at least \SI{1500}{\feet\per\minute}
	\item SDES1500: Strengthen Descent to at least \SI{1500}{\feet\per\minute}
	\item SCL1500: Strengthen Climb to at least \SI{1500}{\feet\per\minute}
	\item SDES2500: Strengthen Descent to at least \SI{2500}{\feet\per\minute}
	\item SCL2500: Strengthen Climb to at least \SI{2500}{\feet\per\minute}
\end{enumerate}

Each advisory instructs the pilot to accelerate until complying with the specified climb or descent rate, except for COC, which allows the pilot freedom to choose any acceleration $\ddot{h}_0 \in [-g/8,g/8]$, where $g$ is the sea-level gravitational acceleration constant, \SI{32.2}{\feet\per\second\squared}. For advisories DNC, DND, DES1500, and CL1500, the pilot is assumed to accelerate in the range $\vert a \vert \in [g/4,g/3]$ with the sign of $\ddot{h}_0$ determined by the specific advisory. If the pilot is already compliant with the given advisory, then the pilot is assumed to continue at the current climbrate. For advisories SDES1500, SCL1500, SDES2500, and SCL2500, the pilot as assumed to accelerate at $\pm g/3$ until compliance. For example, a pilot receiving the CL1500 advisory while descending at \SI{-500}{\feet\per\minute} is assumed to begin accelerating upwards with some acceleration between $g/4$ and $g/3$ and then maintaining a constant climbrate upon reaching the \SI{1500}{\feet\per\minute} climbrate.

New advisories $s'_{\text{adv}}$ are given once each second ($\Delta t=1$), and the state variables are updated as
\begin{equation}
\begin{bmatrix}
h \\
\dot{h}_0 \\
\tau \\
s_{\text{adv}} \\
\end{bmatrix}
\leftarrow
\begin{bmatrix}
h - \dot{h}_0 \Delta t - 0.5\ddot{h}_0 \Delta t^2 \\
\dot{h}_0 + \ddot{h}_0\Delta t \\
\tau-\Delta t \\
s'_{\text{adv}}
\end{bmatrix}
\end{equation}

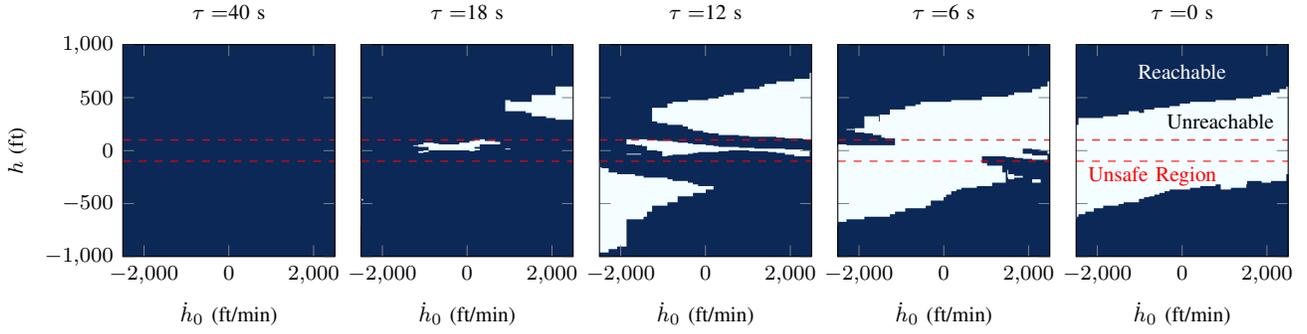
\begin{figure*}
	\centering
	\input{VerticalCAS_ReachableRow.tex}
	\caption{VerticalCAS reachable set over time}
	\label{fig:VC_reachRow}
\end{figure*}

\begin{figure}
	\centering
	\input{VertCAS_Policy.tex}
	\caption{VerticalCAS policy with $\dot{h}_0=-900$ ft/min, $s_{\text{adv}}=\text{DES1500}$}
	\label{fig:VertCAS_policy}
\end{figure}
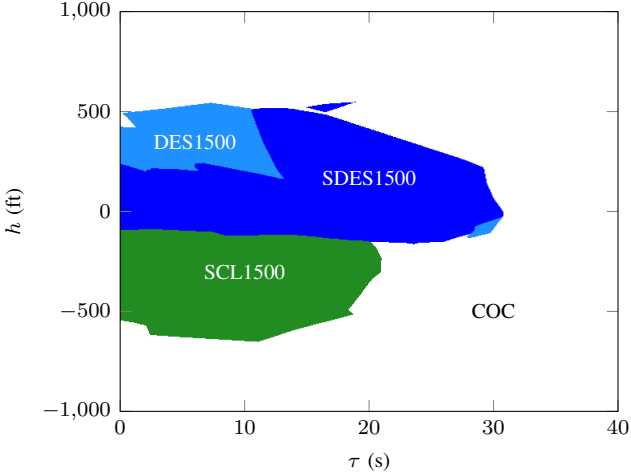

The reward structure contains penalties for NMACs at $\tau=\SI{0}{\second}$, giving non-COC advisories, and reversing/strengthening an advisory. Advisories that transition from a previous weak advisory like COC or DNC to a strong advisory like SCL1500 or SDES2500 are also penalized. The reward structure contains many terms, and Kochenderfer describes in greater detail the reward function used here~\cite{Kochenderfer2015chapter10}. This problem was optimized using discrete value iteration and compressed with a neural network in the same was as described in ~\cref{sec:nnet}. Since $s_{\text{adv}}$ is discrete, nine separate networks were trained, one for each $s_{\text{adv}}$. Each network has three inputs for the three geometric state variables, five ReLU-activated hidden layers of 20 units each, and nine outputs representing the score of each of the nine advisories. VerticalCAS recommends the highest scoring action to the pilot. An example policy plot is shown in \cref{fig:VertCAS_policy}, which shows that the system tends to strengthen the previous advisory in order to avoid collisions. However, if the intruder is well below the aircraft, then descending could result in an NMAC, so the system reverses the advisory and strengthens to SCL1500.

\subsection{Reachability Results}
To define $\mathcal{S}_N$, we first note that $s_{\text{adv}}$ is discrete with nine values, and $\tau$ acts independently and can be discretized to $\{0,1,\dots,40\}$. The remaining variables, $h$ and $\dot{h}_0$, are continuous and need to be discretized to define cells. Because we are interested in the NMAC region where $\vert h \vert \leq \SI{100}{\feet}$, low magnitude $h$ values were more finely discretized to limit over-approximation errors in that region of the state space. The final discretization used 231 points for $h$ and 44 for $\dot{h}_0$, resulting in 3.65 million discrete cells. Reluplex was used to approximate the neural networks, generating 32.8 million queries in order to check for each of the nine advisories in all discrete cells. Reluplex took 175 CPU-hours to complete, an order of magnitude longer than for the mountain car example. 

After approximating the neural networks, the reachable sets can be computed. The reachable region $R_S$ was over-approximated by computing the minimum and maximum $h$ and $\dot{h}_0$ values achievable from cell $s$ given the acceleration constraints. The initial reachable set is all cells at $\tau=\SI{40}{\second}$ with $s_{adv}=\text{COC}$. Each new reachable set counts down $\tau$ until eventually $\tau=\SI{0}{\second}$ and the aircraft must be separated by $\SI{100}{\feet}$ vertically to avoid an NMAC. As seen in \cref{fig:VC_reachRow}, the neural network controller causes the ownship to climb or descend in order to avoid the unsafe NMAC set. This method guarantees that no collision will occur when using the neural network collision avoidance system, given that assumptions made to constrain the acceleration hold. If the acceleration limits are relaxed such that the pilot can accelerate up to $g/2$, for example, then the reachable set will overlap with the NMAC region, and safety cannot be guaranteed.

\begin{comment}
\begin{table}
	\centering
	\caption{VerticalCAS neural network approximation results}
	\begin{tabular}{lrrr}  
		\toprule
		Method  & Num. Actions & Runtime (s) & Complete \\
		\midrule
		Sampling     &  3803229  & 40712  & No  \\
		ReluVal      &  ReluVal ERRORS    & RELUVAL ERRORS & No  \\
		Reluplex     &  3803515  & 630060 & Yes \\
		
		\bottomrule
	\end{tabular}
	\label{tab:VCtable}
\end{table}
\end{comment}

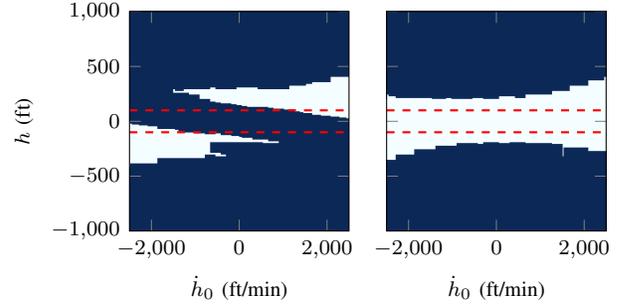
\begin{figure}
	\centering
	\input{VerticalCAS_ReachablePlot.tex}
	\caption{Reachable sets for $\tau=\SI{0}{\second}$, $\epsilon=\SI{3}{\second}$ without (left) and with (right) an online cost to prevent multiple reversals}
	\label{fig:VC_reach}
\end{figure}

\subsection{Pilot Delay}
Assuming that a human pilot will react immediately to collision avoidance advisories from the neural network system is unreasonable. In reality, there will be some amount of pilot delay when following the advisories. For maximum pilot delay $\epsilon$, the pilot could choose to follow any of the advisories of the previous $\epsilon$ seconds. Pilot delay can be incorporated into the reachable set formulation by tracking the $\epsilon$ most recent advisories that result in reaching each cell. A union of cells reachable using all possible delayed advisories is used to compute $\mathcal{R}_{t+1}$ instead of using the current advisory.

For a pilot delay of three seconds, safety is no longer guaranteed, as seen in the left plot of \cref{fig:VC_reach}. For example, consider an ownship slightly below the intruder. VerticalCAS would alert the ownship to descend to avoid an NMAC, but with pilot delay, the ownship could climb for $\epsilon$ seconds instead. At this point, the ownship could be above the intruder aircraft in a region where VerticalCAS would reverse the advisory to make the ownship continue climbing and gain further separation. However, with pilot delay, the pilot could begin following the stale descend advisories and descend instead. The cycle repeats indefinitely, resulting in an NMAC.

One solution to fix such issues uses online costs to penalize undesirable behavior~\cite{Kochenderfer2015chapter10}. For example, an online cost could be applied to prevent the system from reversing the direction of given advisories more than once. A count of reversals can be tracked along with the previous advisories given to reach each cell. As shown in \Cref{fig:VC_reach}, preventing double reversals is sufficient to prevent an NMAC. This reachability method allows complex systems involving neural network controllers alongside other variables and penalties to be analyzed to ensure safety.

\section{Conclusions}

Guaranteeing the performance of neural network controllers in the presence of uncertainty is paramount to incorporating such controllers in safety critical systems. We presented a general method that uses existing neural network verification tools to constrain the neural network output before computing over-approximated reachable sets of system states. If the over-approximated set does not contain any unsafe states, then the system is guaranteed to be safe. A mountain car example demonstrated the approach and provided guarantees on time to reach the goal for different levels of disturbances. A collision avoidance example demonstrated how complex systems involving human delay and online costs could be modeled and shown to be safe. The proposed method is flexible and applicable to any neural network controller of a dynamical system. Future work will explore higher dimensional problems along with open source reachable set libraries in order to decrease computation time.

\section*{Acknowledgments}
This material is based upon work supported by the National Science Foundation Graduate Research Fellowship under Grant No. DGE$-$1656518. Any opinion, findings, and conclusions or recommendations expressed in this material are those of the authors and do not necessarily reflect the views of the National Science Foundation.

%\bibliographystyle{IEEEtran}
%\bibliography{references}

\printbibliography

\end{document}

%% file: MountainCarPolicy.tex
\begin{tikzpicture}[]
\begin{groupplot}[group style={horizontal sep=4cm, vertical sep=2.5cm, group size=2 by 2}]
\nextgroupplot [height = {7.0cm}, ylabel = {Speed}, xmin = {-1.2}, xmax = {0.6}, ymax = {0.07}, xlabel = {Position}, ymin = {-0.07}, width = {8.6cm}, enlargelimits = false, axis on top]\addplot [point meta min=1, point meta max=3] graphics [xmin=-1.2, xmax=0.6, ymin=-0.07, ymax=0.07] {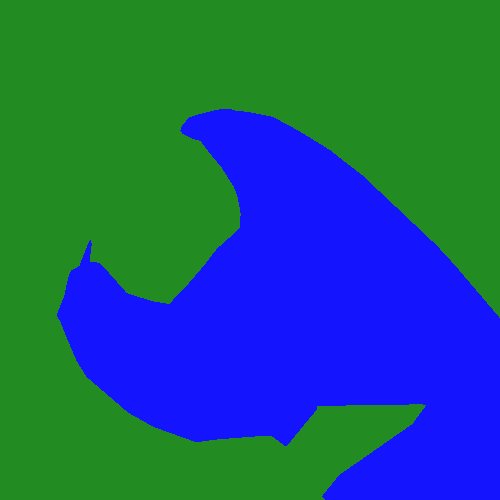};
\node[text=white] at (0.2,0.05)  {$u=1$ (Right)};
\node[text=white] at (-0.1,-0.015)  {$u=-1$ (Left)};
\end{groupplot}

\end{tikzpicture}

%% file: GridDensity.tex
\begin{tikzpicture}[]
\begin{axis}[height = {7cm}, ylabel = {Speed}, xlabel = {Position}, width = {7.2cm}, enlargelimits = false, axis on top, colormap={mycolormap}{ rgb(0cm)=(0.0,0.0,0.515625) rgb(1cm)=(0.0,0.0,0.53125) rgb(2cm)=(0.0,0.0,0.546875) rgb(3cm)=(0.0,0.0,0.5625) rgb(4cm)=(0.0,0.0,0.578125) rgb(5cm)=(0.0,0.0,0.59375) rgb(6cm)=(0.0,0.0,0.609375) rgb(7cm)=(0.0,0.0,0.625) rgb(8cm)=(0.0,0.0,0.640625) rgb(9cm)=(0.0,0.0,0.65625) rgb(10cm)=(0.0,0.0,0.671875) rgb(11cm)=(0.0,0.0,0.6875) rgb(12cm)=(0.0,0.0,0.703125) rgb(13cm)=(0.0,0.0,0.71875) rgb(14cm)=(0.0,0.0,0.734375) rgb(15cm)=(0.0,0.0,0.75) rgb(16cm)=(0.0,0.0,0.765625) rgb(17cm)=(0.0,0.0,0.78125) rgb(18cm)=(0.0,0.0,0.796875) rgb(19cm)=(0.0,0.0,0.8125) rgb(20cm)=(0.0,0.0,0.828125) rgb(21cm)=(0.0,0.0,0.84375) rgb(22cm)=(0.0,0.0,0.859375) rgb(23cm)=(0.0,0.0,0.875) rgb(24cm)=(0.0,0.0,0.890625) rgb(25cm)=(0.0,0.0,0.90625) rgb(26cm)=(0.0,0.0,0.921875) rgb(27cm)=(0.0,0.0,0.9375) rgb(28cm)=(0.0,0.0,0.953125) rgb(29cm)=(0.0,0.0,0.96875) rgb(30cm)=(0.0,0.0,0.984375) rgb(31cm)=(0.0,0.0,1.0) rgb(32cm)=(0.0,0.015625,1.0) rgb(33cm)=(0.0,0.03125,1.0) rgb(34cm)=(0.0,0.046875,1.0) rgb(35cm)=(0.0,0.0625,1.0) rgb(36cm)=(0.0,0.078125,1.0) rgb(37cm)=(0.0,0.09375,1.0) rgb(38cm)=(0.0,0.109375,1.0) rgb(39cm)=(0.0,0.125,1.0) rgb(40cm)=(0.0,0.140625,1.0) rgb(41cm)=(0.0,0.15625,1.0) rgb(42cm)=(0.0,0.171875,1.0) rgb(43cm)=(0.0,0.1875,1.0) rgb(44cm)=(0.0,0.203125,1.0) rgb(45cm)=(0.0,0.21875,1.0) rgb(46cm)=(0.0,0.234375,1.0) rgb(47cm)=(0.0,0.25,1.0) rgb(48cm)=(0.0,0.265625,1.0) rgb(49cm)=(0.0,0.28125,1.0) rgb(50cm)=(0.0,0.296875,1.0) rgb(51cm)=(0.0,0.3125,1.0) rgb(52cm)=(0.0,0.328125,1.0) rgb(53cm)=(0.0,0.34375,1.0) rgb(54cm)=(0.0,0.359375,1.0) rgb(55cm)=(0.0,0.375,1.0) rgb(56cm)=(0.0,0.390625,1.0) rgb(57cm)=(0.0,0.40625,1.0) rgb(58cm)=(0.0,0.421875,1.0) rgb(59cm)=(0.0,0.4375,1.0) rgb(60cm)=(0.0,0.453125,1.0) rgb(61cm)=(0.0,0.46875,1.0) rgb(62cm)=(0.0,0.484375,1.0) rgb(63cm)=(0.0,0.5,1.0) rgb(64cm)=(0.0,0.515625,1.0) rgb(65cm)=(0.0,0.53125,1.0) rgb(66cm)=(0.0,0.546875,1.0) rgb(67cm)=(0.0,0.5625,1.0) rgb(68cm)=(0.0,0.578125,1.0) rgb(69cm)=(0.0,0.59375,1.0) rgb(70cm)=(0.0,0.609375,1.0) rgb(71cm)=(0.0,0.625,1.0) rgb(72cm)=(0.0,0.640625,1.0) rgb(73cm)=(0.0,0.65625,1.0) rgb(74cm)=(0.0,0.671875,1.0) rgb(75cm)=(0.0,0.6875,1.0) rgb(76cm)=(0.0,0.703125,1.0) rgb(77cm)=(0.0,0.71875,1.0) rgb(78cm)=(0.0,0.734375,1.0) rgb(79cm)=(0.0,0.75,1.0) rgb(80cm)=(0.0,0.765625,1.0) rgb(81cm)=(0.0,0.78125,1.0) rgb(82cm)=(0.0,0.796875,1.0) rgb(83cm)=(0.0,0.8125,1.0) rgb(84cm)=(0.0,0.828125,1.0) rgb(85cm)=(0.0,0.84375,1.0) rgb(86cm)=(0.0,0.859375,1.0) rgb(87cm)=(0.0,0.875,1.0) rgb(88cm)=(0.0,0.890625,1.0) rgb(89cm)=(0.0,0.90625,1.0) rgb(90cm)=(0.0,0.921875,1.0) rgb(91cm)=(0.0,0.9375,1.0) rgb(92cm)=(0.0,0.953125,1.0) rgb(93cm)=(0.0,0.96875,1.0) rgb(94cm)=(0.0,0.984375,1.0) rgb(95cm)=(0.0,1.0,1.0) rgb(96cm)=(0.015625,1.0,0.984375) rgb(97cm)=(0.03125,1.0,0.96875) rgb(98cm)=(0.046875,1.0,0.953125) rgb(99cm)=(0.0625,1.0,0.9375) rgb(100cm)=(0.078125,1.0,0.921875) rgb(101cm)=(0.09375,1.0,0.90625) rgb(102cm)=(0.109375,1.0,0.890625) rgb(103cm)=(0.125,1.0,0.875) rgb(104cm)=(0.140625,1.0,0.859375) rgb(105cm)=(0.15625,1.0,0.84375) rgb(106cm)=(0.171875,1.0,0.828125) rgb(107cm)=(0.1875,1.0,0.8125) rgb(108cm)=(0.203125,1.0,0.796875) rgb(109cm)=(0.21875,1.0,0.78125) rgb(110cm)=(0.234375,1.0,0.765625) rgb(111cm)=(0.25,1.0,0.75) rgb(112cm)=(0.265625,1.0,0.734375) rgb(113cm)=(0.28125,1.0,0.71875) rgb(114cm)=(0.296875,1.0,0.703125) rgb(115cm)=(0.3125,1.0,0.6875) rgb(116cm)=(0.328125,1.0,0.671875) rgb(117cm)=(0.34375,1.0,0.65625) rgb(118cm)=(0.359375,1.0,0.640625) rgb(119cm)=(0.375,1.0,0.625) rgb(120cm)=(0.390625,1.0,0.609375) rgb(121cm)=(0.40625,1.0,0.59375) rgb(122cm)=(0.421875,1.0,0.578125) rgb(123cm)=(0.4375,1.0,0.5625) rgb(124cm)=(0.453125,1.0,0.546875) rgb(125cm)=(0.46875,1.0,0.53125) rgb(126cm)=(0.484375,1.0,0.515625) rgb(127cm)=(0.5,1.0,0.5) rgb(128cm)=(0.515625,1.0,0.484375) rgb(129cm)=(0.53125,1.0,0.46875) rgb(130cm)=(0.546875,1.0,0.453125) rgb(131cm)=(0.5625,1.0,0.4375) rgb(132cm)=(0.578125,1.0,0.421875) rgb(133cm)=(0.59375,1.0,0.40625) rgb(134cm)=(0.609375,1.0,0.390625) rgb(135cm)=(0.625,1.0,0.375) rgb(136cm)=(0.640625,1.0,0.359375) rgb(137cm)=(0.65625,1.0,0.34375) rgb(138cm)=(0.671875,1.0,0.328125) rgb(139cm)=(0.6875,1.0,0.3125) rgb(140cm)=(0.703125,1.0,0.296875) rgb(141cm)=(0.71875,1.0,0.28125) rgb(142cm)=(0.734375,1.0,0.265625) rgb(143cm)=(0.75,1.0,0.25) rgb(144cm)=(0.765625,1.0,0.234375) rgb(145cm)=(0.78125,1.0,0.21875) rgb(146cm)=(0.796875,1.0,0.203125) rgb(147cm)=(0.8125,1.0,0.1875) rgb(148cm)=(0.828125,1.0,0.171875) rgb(149cm)=(0.84375,1.0,0.15625) rgb(150cm)=(0.859375,1.0,0.140625) rgb(151cm)=(0.875,1.0,0.125) rgb(152cm)=(0.890625,1.0,0.109375) rgb(153cm)=(0.90625,1.0,0.09375) rgb(154cm)=(0.921875,1.0,0.078125) rgb(155cm)=(0.9375,1.0,0.0625) rgb(156cm)=(0.953125,1.0,0.046875) rgb(157cm)=(0.96875,1.0,0.03125) rgb(158cm)=(0.984375,1.0,0.015625) rgb(159cm)=(1.0,1.0,0.0) rgb(160cm)=(1.0,0.984375,0.0) rgb(161cm)=(1.0,0.96875,0.0) rgb(162cm)=(1.0,0.953125,0.0) rgb(163cm)=(1.0,0.9375,0.0) rgb(164cm)=(1.0,0.921875,0.0) rgb(165cm)=(1.0,0.90625,0.0) rgb(166cm)=(1.0,0.890625,0.0) rgb(167cm)=(1.0,0.875,0.0) rgb(168cm)=(1.0,0.859375,0.0) rgb(169cm)=(1.0,0.84375,0.0) rgb(170cm)=(1.0,0.828125,0.0) rgb(171cm)=(1.0,0.8125,0.0) rgb(172cm)=(1.0,0.796875,0.0) rgb(173cm)=(1.0,0.78125,0.0) rgb(174cm)=(1.0,0.765625,0.0) rgb(175cm)=(1.0,0.75,0.0) rgb(176cm)=(1.0,0.734375,0.0) rgb(177cm)=(1.0,0.71875,0.0) rgb(178cm)=(1.0,0.703125,0.0) rgb(179cm)=(1.0,0.6875,0.0) rgb(180cm)=(1.0,0.671875,0.0) rgb(181cm)=(1.0,0.65625,0.0) rgb(182cm)=(1.0,0.640625,0.0) rgb(183cm)=(1.0,0.625,0.0) rgb(184cm)=(1.0,0.609375,0.0) rgb(185cm)=(1.0,0.59375,0.0) rgb(186cm)=(1.0,0.578125,0.0) rgb(187cm)=(1.0,0.5625,0.0) rgb(188cm)=(1.0,0.546875,0.0) rgb(189cm)=(1.0,0.53125,0.0) rgb(190cm)=(1.0,0.515625,0.0) rgb(191cm)=(1.0,0.5,0.0) rgb(192cm)=(1.0,0.484375,0.0) rgb(193cm)=(1.0,0.46875,0.0) rgb(194cm)=(1.0,0.453125,0.0) rgb(195cm)=(1.0,0.4375,0.0) rgb(196cm)=(1.0,0.421875,0.0) rgb(197cm)=(1.0,0.40625,0.0) rgb(198cm)=(1.0,0.390625,0.0) rgb(199cm)=(1.0,0.375,0.0) rgb(200cm)=(1.0,0.359375,0.0) rgb(201cm)=(1.0,0.34375,0.0) rgb(202cm)=(1.0,0.328125,0.0) rgb(203cm)=(1.0,0.3125,0.0) rgb(204cm)=(1.0,0.296875,0.0) rgb(205cm)=(1.0,0.28125,0.0) rgb(206cm)=(1.0,0.265625,0.0) rgb(207cm)=(1.0,0.25,0.0) rgb(208cm)=(1.0,0.234375,0.0) rgb(209cm)=(1.0,0.21875,0.0) rgb(210cm)=(1.0,0.203125,0.0) rgb(211cm)=(1.0,0.1875,0.0) rgb(212cm)=(1.0,0.171875,0.0) rgb(213cm)=(1.0,0.15625,0.0) rgb(214cm)=(1.0,0.140625,0.0) rgb(215cm)=(1.0,0.125,0.0) rgb(216cm)=(1.0,0.109375,0.0) rgb(217cm)=(1.0,0.09375,0.0) rgb(218cm)=(1.0,0.078125,0.0) rgb(219cm)=(1.0,0.0625,0.0) rgb(220cm)=(1.0,0.046875,0.0) rgb(221cm)=(1.0,0.03125,0.0) rgb(222cm)=(1.0,0.015625,0.0) rgb(223cm)=(1.0,0.0,0.0) rgb(224cm)=(0.984375,0.0,0.0) rgb(225cm)=(0.96875,0.0,0.0) rgb(226cm)=(0.953125,0.0,0.0) rgb(227cm)=(0.9375,0.0,0.0) rgb(228cm)=(0.921875,0.0,0.0) rgb(229cm)=(0.90625,0.0,0.0) rgb(230cm)=(0.890625,0.0,0.0) rgb(231cm)=(0.875,0.0,0.0) rgb(232cm)=(0.859375,0.0,0.0) rgb(233cm)=(0.84375,0.0,0.0) rgb(234cm)=(0.828125,0.0,0.0) rgb(235cm)=(0.8125,0.0,0.0) rgb(236cm)=(0.796875,0.0,0.0) rgb(237cm)=(0.78125,0.0,0.0) rgb(238cm)=(0.765625,0.0,0.0) rgb(239cm)=(0.75,0.0,0.0) rgb(240cm)=(0.734375,0.0,0.0) rgb(241cm)=(0.71875,0.0,0.0) rgb(242cm)=(0.703125,0.0,0.0) rgb(243cm)=(0.6875,0.0,0.0) rgb(244cm)=(0.671875,0.0,0.0) rgb(245cm)=(0.65625,0.0,0.0) rgb(246cm)=(0.640625,0.0,0.0) rgb(247cm)=(0.625,0.0,0.0) rgb(248cm)=(0.609375,0.0,0.0) rgb(249cm)=(0.59375,0.0,0.0) rgb(250cm)=(0.578125,0.0,0.0) rgb(251cm)=(0.5625,0.0,0.0) rgb(252cm)=(0.546875,0.0,0.0) rgb(253cm)=(0.53125,0.0,0.0) rgb(254cm)=(0.515625,0.0,0.0) }, colorbar, colorbar style = {ylabel={Number of Nearby Cells (Thousands)}}]\addplot [point meta min=0.0, point meta max=5.0] graphics [xmin=-1.2, xmax=0.6, ymin=-0.07, ymax=0.07] {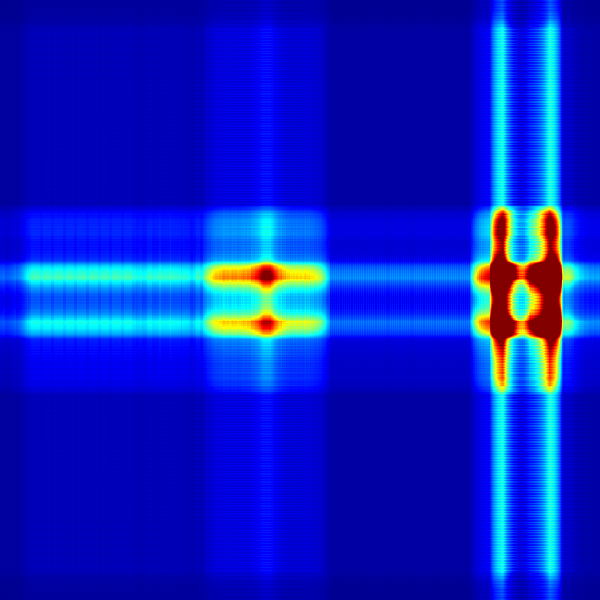};
\end{axis}

\end{tikzpicture}

%% file: MC_ReachableRow.tex
\begin{tikzpicture}[]
\begin{groupplot}[height=4.0cm, width=4.0cm, group style={horizontal sep=0.35cm, group size=6 by 1}]
\nextgroupplot [ylabel = {Velocity}, title = {Time Steps: 0}, xlabel = {Position}, , enlargelimits = false, axis on top]\addplot [point meta min=0, point meta max=1] graphics [xmin=-1.2, xmax=0.6, ymin=-0.07, ymax=0.07] {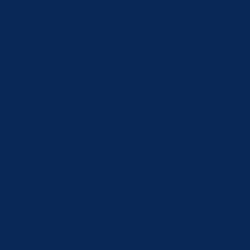};
\nextgroupplot [ylabel = {}, title = {Time Steps: 10}, xlabel = {Position}, yticklabels={,,}, scaled y ticks=false, enlargelimits = false, axis on top]\addplot [point meta min=0, point meta max=1] graphics [xmin=-1.2, xmax=0.6, ymin=-0.07, ymax=0.07] {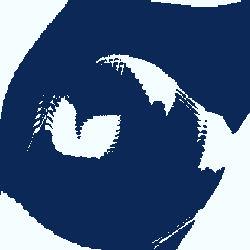};
\nextgroupplot [ylabel = {}, title = {Time Steps: 40}, xlabel = {Position}, yticklabels={,,}, scaled y ticks=false, enlargelimits = false, axis on top]\addplot [point meta min=0, point meta max=1] graphics [xmin=-1.2, xmax=0.6, ymin=-0.07, ymax=0.07] {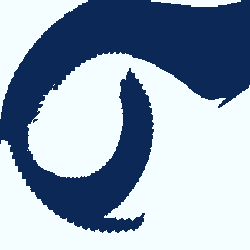};
\nextgroupplot [ylabel = {}, title = {Time Steps: 80}, xlabel = {Position}, yticklabels={,,}, scaled y ticks=false, enlargelimits = false, axis on top]\addplot [point meta min=0, point meta max=1] graphics [xmin=-1.2, xmax=0.6, ymin=-0.07, ymax=0.07] {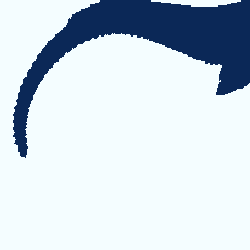};

\node[text=white] at (0,0.058)  {\footnotesize Reachable};
\node[text=black] at (-0.2,-0.03)  {\footnotesize Unreachable};

\nextgroupplot [ylabel = {}, title = {Time Steps: 100}, xlabel = {Position}, yticklabels={,,}, scaled y ticks=false, enlargelimits = false, axis on top]\addplot [point meta min=0, point meta max=1] graphics [xmin=-1.2, xmax=0.6, ymin=-0.07, ymax=0.07] {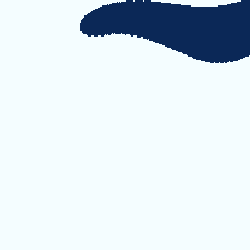};
\nextgroupplot [ylabel = {}, title = {Time Steps: 130}, xlabel = {Position}, yticklabels={,,}, scaled y ticks=false, enlargelimits = false, axis on top]\addplot [point meta min=0, point meta max=1] graphics [xmin=-1.2, xmax=0.6, ymin=-0.07, ymax=0.07] {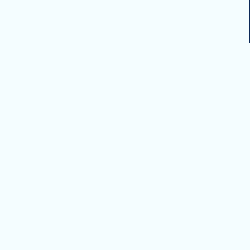};

\end{groupplot}

\end{tikzpicture}

%% file: ReachableCells.tex
\begin{tikzpicture}[]
\begin{axis}[height={7cm},legend pos = {south west}, ylabel = {Velocity}, xlabel = {Position},
x tick label style={
	/pgf/number format/.cd,
	fixed,
	precision=3,
	/tikz/.cd
}]
\addplot+ [mark = {*}, black, thick, solid]coordinates {
	(0.22235185185185186, 0.021338636363636365)
	(0.22789814814814816, 0.021338636363636365)
	(0.22789814814814816, 0.0219625)
	(0.22235185185185186, 0.0219625)
	(0.22235185185185186, 0.021338636363636365)
};
\addlegendentry{$S$}
\addplot+ [thick]coordinates {
	(0.24923678451178452, 0.020124519553109586)
	(0.24986064814814815, 0.020774389058416626)
	(0.24986064814814815, 0.022274389058416624)
	(0.24431435185185185, 0.022274389058416624)
	(0.24369048821548822, 0.021650525422052994)
	(0.24369048821548822, 0.020124519553109586)
	(0.24923678451178452, 0.020124519553109586)
};
\addlegendentry{$R_S$}
\addplot+ [mark = {diamond*}, mark size = {3}, cyan,thick,solid,mark options={fill=cyan}]coordinates {
	(0.23899074074074075, 0.02009090909090909)
	(0.24453703703703702, 0.02009090909090909)
	(0.24453703703703702, 0.020714772727272728)
	(0.23899074074074075, 0.020714772727272728)
	(0.23899074074074075, 0.02009090909090909)
};
\addlegendentry{$\mathcal{R}^S$}
\addplot+[draw=none, mark options={fill=green!50!black,draw=green!50!black},only marks,mark size=2,mark=*] coordinates {
	(0.22740436357362054, 0.02194474764705569)
};
\addlegendentry{Random samples}
\addplot+[draw=none, mark options={fill=blue,draw=blue},only marks,mark size=2,mark=*] coordinates {
	(0.22740436357362054, 0.02194474764705569)
};
\addlegendentry{Updated samples}
\addplot+[draw=none, mark options={fill=green!50!black,draw=green!50!black},only marks,mark size=2,mark=*] coordinates {
	(0.22740436357362054, 0.02194474764705569)
};
\addplot+[draw=none, mark options={fill=blue,draw=blue},only marks,mark size=1,mark=*] coordinates {
	(0.24934911122067624, 0.022016677505265807)
	(0.24836661338473143, 0.021089392375580475)
	(0.24559556427438028, 0.021649888786642972)
	(0.24554952119710668, 0.020256812357412796)
	(0.24600844680079742, 0.021485062130116767)
	(0.246006293845179, 0.021076638397475485)
	(0.2474660981362513, 0.02178312500181044)
	(0.2486437112804635, 0.021505803402703076)
	(0.24954694993233037, 0.020850602164438632)
	(0.24820632069991572, 0.020856972483343256)
	(0.24676244628277397, 0.02209790377172268)
	(0.24446747872651203, 0.0210033737141367)
	(0.24631491461661711, 0.020556980771640237)
	(0.2481783558174043, 0.02020591664531829)
	(0.24827878398808972, 0.020253319572553295)
	(0.2473446559505051, 0.02181576667806439)
	(0.2457198970439919, 0.02045493775041478)
	(0.24908318784737993, 0.020883482816927007)
	(0.24454655269818473, 0.021479432954165777)
	(0.24414702345150135, 0.020947859640718526)
	(0.24932595592752765, 0.020601379991897897)
	(0.24686424224109643, 0.02079698702243551)
	(0.24764890745339735, 0.021166764102814924)
	(0.24582290472304, 0.020877316351964863)
	(0.2470785112032927, 0.020370625753874105)
	(0.2497793017695652, 0.021924407802663026)
	(0.2442852074001492, 0.021566017175937056)
	(0.24686626886870217, 0.021038220005301308)
	(0.24670367379715952, 0.02088917910652393)
	(0.24880327623299725, 0.021926472300863482)
	(0.2442471247060432, 0.021244827638871335)
	(0.24459188613985894, 0.021581864754109007)
	(0.24524158232453364, 0.02099366001313141)
	(0.2487615329145172, 0.021139452080260687)
	(0.24469970219877832, 0.02099406898658241)
	(0.24730516282399762, 0.02193209537788733)
	(0.24551626302488885, 0.02177011339097039)
	(0.24827556403872814, 0.0213677190501906)
	(0.24822926936373524, 0.02062212546093417)
	(0.24799913120245604, 0.020871421408323453)
	(0.24483897677757496, 0.02149160447535033)
	(0.24592904663525195, 0.021158492704795)
	(0.24797614055391723, 0.02162480337860904)
	(0.24784574548135369, 0.02095968166832728)
	(0.24482675956403457, 0.02193527928384866)
	(0.24645018859335, 0.02169285465899198)
	(0.24496951407132567, 0.021775322127261652)
	(0.24455065281734334, 0.02103568258362401)
	(0.24653407238825775, 0.02104849186311085)
	(0.24775493481828814, 0.02074243878090854)
	(0.24830452095963573, 0.021676019087774754)
	(0.24748082862543536, 0.021782616014558772)
	(0.24792034469026164, 0.02140577309891499)
	(0.2454134282763825, 0.021903003948435196)
	(0.2448018783390898, 0.021319596209280707)
	(0.24560902563103196, 0.021158007521308652)
	(0.24635820696027544, 0.02061386478852961)
	(0.2476565438289589, 0.020579359090336695)
	(0.24776052018330086, 0.02185415822724027)
	(0.24575703043700808, 0.02078711715495293)
	(0.24489039830403647, 0.02053665050298236)
	(0.24677217115521857, 0.021352241135059263)
	(0.24838787215521713, 0.021067912383690542)
	(0.24756370990758564, 0.02127713163464912)
	(0.24762228476273881, 0.02112313103997667)
	(0.24668222385016914, 0.021343388957905822)
	(0.24866749204918823, 0.0212968023150404)
	(0.24519682833784526, 0.02130941017557694)
	(0.2444700122123138, 0.0219132274859154)
	(0.24403146593152814, 0.020237690753556144)
	(0.24719000466778412, 0.021290236085667755)
	(0.2491336929505505, 0.020723193672179535)
	(0.2448837925817362, 0.02138889767133944)
	(0.2472853402627031, 0.022003671290841326)
	(0.24415011020866936, 0.021118386077247135)
	(0.2476992683677988, 0.02128139744358255)
	(0.24713093148034473, 0.020845116848255716)
	(0.2463318820727574, 0.022118769575816315)
	(0.2442120547955227, 0.021050867870432866)
	(0.24552294261605237, 0.02093842674814976)
	(0.2478898806416883, 0.021521997786288466)
	(0.24894662641247414, 0.021321249804577298)
	(0.24828977618044726, 0.020844531035246413)
	(0.24892107183644407, 0.02148664740750279)
	(0.24860993267360293, 0.0211883756631073)
	(0.24570448108317644, 0.020599457430701153)
	(0.2457494871056483, 0.02152775101452754)
	(0.24702978241406487, 0.020764419922749425)
	(0.24747213116849706, 0.020575043536297705)
	(0.24648307965181912, 0.02142130305528564)
	(0.24841007950720914, 0.02066639882415592)
	(0.2494975613089607, 0.020893714692300357)
	(0.2469624259522562, 0.021662714875975327)
	(0.24652993912265525, 0.021816152746028238)
	(0.24480811158701887, 0.02073829843924884)
	(0.24783961844641267, 0.02096821585722225)
	(0.2459234350205771, 0.021001911464087433)
	(0.24846751628829905, 0.021720994501747394)
	(0.24749125675821337, 0.022078125796684316)
	(0.24755950153465706, 0.02176098237312634)
	(0.24489699068578366, 0.020731151081474945)
	(0.24516912527129622, 0.021688035432070752)
	(0.24666472677068205, 0.021852570765819458)
	(0.2448422878075389, 0.02055346089722878)
	(0.2461697621059808, 0.021132500843229755)
	(0.24908437883276216, 0.02130360399921413)
	(0.24657044903231312, 0.020604875123298574)
	(0.2472634659466065, 0.02115190491012153)
	(0.2464980681165381, 0.0221368292213752)
	(0.24393520566217852, 0.020519227016886736)
	(0.2483222602075644, 0.020397984054869478)
	(0.24645648060869016, 0.021600076568202103)
	(0.247491732469782, 0.021037018286743685)
	(0.24734297007288783, 0.021026490491718824)
	(0.24530842660447263, 0.02151865797221866)
	(0.24448796496169534, 0.02116852887203801)
	(0.24725172205519297, 0.021977957174322008)
	(0.2471903677882828, 0.021314800383496153)
	(0.24592997254861307, 0.020609273661704012)
	(0.24460385840105708, 0.021333709836479024)
	(0.24515680112774976, 0.020872231270643384)
	(0.2471747608915262, 0.020739060096747454)
	(0.2472669258140818, 0.02027875663856636)
	(0.2442666857079704, 0.021932775744765445)
	(0.24950459849465997, 0.020868079316590786)
	(0.2490725554318274, 0.021536144533612772)
	(0.24803471946160915, 0.02167925430837393)
	(0.24407479311770883, 0.02077330655465661)
	(0.24573969381120497, 0.021327668204178356)
	(0.2478546652199351, 0.021770939408832096)
	(0.2453677399034513, 0.020863982805135312)
	(0.24597178531862351, 0.020656631326144946)
	(0.2441274515324931, 0.020935247643047775)
	(0.2473516211603611, 0.021536896491839822)
	(0.24803454399313757, 0.02166832930149413)
	(0.24600468429085665, 0.02165384786971422)
	(0.24644517244418568, 0.02194178128580769)
	(0.24579330648431166, 0.02065961580715492)
	(0.24499910667021654, 0.021885950706894808)
	(0.24796623371555787, 0.021627491305053623)
	(0.2460826292226621, 0.020466487456685768)
	(0.2464503918386658, 0.020495669553767085)
	(0.24651844178542628, 0.020740170386545236)
	(0.24761336246440308, 0.021223858409358774)
	(0.24731424264960314, 0.020343154397162697)
	(0.24852828697343324, 0.02085523761117013)
	(0.24594915394276845, 0.021185073030785935)
	(0.24791987258307524, 0.02165038527404222)
	(0.24568146414899902, 0.0221082096028819)
	(0.24750768495444964, 0.021742106563119708)
	(0.24817857662759396, 0.021857490716461484)
	(0.24893533639753324, 0.021554628658428797)
	(0.2471596450699305, 0.020748263198732853)
	(0.2478987744584054, 0.021643039666845592)
	(0.24921402587605637, 0.021034833966661302)
	(0.2458173606989704, 0.020694424042140096)
	(0.24739171328927634, 0.020999714780882417)
	(0.24622522337588984, 0.020601516496394502)
	(0.2475473387692375, 0.022120708685038626)
	(0.24552053425293685, 0.020935179209465637)
	(0.24832784665599478, 0.021036552139773318)
	(0.24781938570888257, 0.02175375774500408)
	(0.24942843802470396, 0.021795140671983376)
	(0.24767410179959465, 0.021440576337234993)
	(0.2469598004410287, 0.02155133363720503)
	(0.24702681058463338, 0.02169711712376048)
	(0.24947524172893615, 0.020650283792635088)
	(0.24433651777393386, 0.022139035141192673)
	(0.2493257440029108, 0.020648665216261606)
	(0.24825356332805645, 0.0203375412094007)
	(0.24752612520468795, 0.02091726279726563)
	(0.2467520966062341, 0.021339589139621952)
	(0.24517456589760891, 0.021097047209132545)
	(0.24742908743483807, 0.020871514317847263)
	(0.2459508226877138, 0.021149973769628617)
	(0.2468126954486122, 0.02107558956322997)
	(0.24781193871392038, 0.021725741947207725)
	(0.24522161067016238, 0.02142867342617112)
	(0.24426115779345886, 0.021403698400983745)
	(0.24820150739345154, 0.021931731689792507)
	(0.2489916972606723, 0.021628438194043413)
	(0.2492563217213304, 0.020914341233826694)
	(0.245520457753996, 0.02143285765701521)
	(0.24627302196407133, 0.021127046096126312)
	(0.24610273665171226, 0.02186854554098141)
	(0.24908300355401156, 0.021045495953009336)
	(0.2473615948708132, 0.020350834081951975)
	(0.24789899373027002, 0.021317471337098978)
	(0.24436985489178742, 0.020769923448489103)
	(0.2464493458686305, 0.020900995657377477)
	(0.2496073971054528, 0.021162147618893984)
	(0.248269185912956, 0.02092123505780293)
	(0.24530894021321156, 0.020937357023560702)
	(0.24606948819478786, 0.020757443915974002)
	(0.245784964946801, 0.02123858687612332)
	(0.2478808717948852, 0.020654371107640998)
	(0.2452853479608363, 0.020842858651491887)
	(0.2467748091970093, 0.02135162830586914)
	(0.24668060177686954, 0.020790680647385262)
	(0.24937051611310335, 0.02161701548710232)
	(0.24847519466142837, 0.020458940333199685)
	(0.24488602759386938, 0.021359647177212202)
	(0.24863795674452643, 0.021736151441538683)
	(0.24810456762620187, 0.020685122516605398)
	(0.24865037823687258, 0.021078579127243362)
	(0.2481384844522673, 0.021816019137209076)
	(0.24870821736475007, 0.02080003038218971)
	(0.24512815837232885, 0.021640140867589835)
	(0.24592387708301072, 0.020560795636123126)
	(0.24437185584066987, 0.021043623390568044)
	(0.24500926743091614, 0.021365642750012823)
	(0.2468001518647385, 0.021597474969443115)
	(0.2478477364017805, 0.020626123356068213)
	(0.2451770915166241, 0.021705045288933307)
	(0.24889225152953437, 0.021377655496538642)
	(0.24764981283509807, 0.021232900272807036)
	(0.24586637084364896, 0.020470376731321703)
	(0.24839729644083944, 0.02133374564997368)
	(0.24587207292920849, 0.02168213563602388)
	(0.24708885503066724, 0.021302860887882698)
	(0.2473685939917885, 0.021489114078584493)
	(0.24644030484201215, 0.020741131880998877)
	(0.24832387878705436, 0.021811035663697527)
	(0.24952332960784585, 0.021021664051557854)
	(0.24955629000774823, 0.02162394473133206)
	(0.24440263900500533, 0.021312943464789045)
	(0.2484024845621497, 0.020806683299222885)
	(0.2447870477220758, 0.02206572419591227)
	(0.24403602850148645, 0.02108917197215225)
	(0.24433102884642632, 0.021766753429436424)
	(0.248948715295979, 0.020665807816912126)
	(0.24666103853327614, 0.02205084326847624)
	(0.244578769435407, 0.020909435378661557)
	(0.24843617814296218, 0.02132785850188951)
	(0.24414720558151345, 0.02162507495638477)
	(0.24532278178861347, 0.020391199103529184)
	(0.24604774962322978, 0.020782842647467777)
	(0.24785485908735458, 0.021184279611160418)
	(0.24701203400414662, 0.020560764981193122)
	(0.24830731608016765, 0.021081584059370127)
	(0.24695637241864454, 0.0213382155081613)
	(0.24827124575706708, 0.02136051478259102)
	(0.2474043961953622, 0.020687660243352918)
	(0.24635815479384127, 0.021363409894054296)
	(0.24917373510843205, 0.02123418459716347)
	(0.2472146290397141, 0.0219697999452314)
	(0.24736735893062428, 0.0213443527122704)
	(0.2491606078779147, 0.02056698667925882)
	(0.24586187555862185, 0.02113229637175803)
	(0.24483798201120197, 0.0216966111807319)
	(0.24440756813141065, 0.021193199471126963)
	(0.24541898516115504, 0.020755750920473704)
	(0.2444373967242213, 0.021572849252118546)
	(0.2490064111256924, 0.02081116525851967)
	(0.2442010964065071, 0.020325138555240804)
	(0.2450029709449636, 0.02181192745614936)
	(0.2460496815246926, 0.02133634597175492)
	(0.24448174836059122, 0.02099168411746867)
	(0.24562886664451183, 0.02130369396260744)
	(0.24685926164998426, 0.020670572589409058)
	(0.24540897701605427, 0.020581568562146726)
	(0.24536765898589497, 0.021367762230091843)
	(0.2472075429250165, 0.020616724090869086)
	(0.2478015694820691, 0.02079482155460866)
	(0.2461470701467662, 0.02123114446988792)
	(0.24793053149139982, 0.021864500035806145)
	(0.2471304544714621, 0.020405124631472753)
	(0.24605766737613818, 0.021210767654424367)
	(0.24461321158199462, 0.02121699476256282)
	(0.24672097654897115, 0.021015983149537783)
	(0.24603177201513915, 0.02123831057106143)
	(0.24696705570411664, 0.02053392370283828)
	(0.24690448195440823, 0.021541596903086883)
	(0.24724796200270394, 0.02131474038405921)
	(0.24857090562914463, 0.02199276909333107)
	(0.24475130184664295, 0.02076097067185463)
	(0.24686351951529778, 0.02161679625522028)
	(0.24895875763679676, 0.021521899909977854)
	(0.24435573020274878, 0.021293242619790476)
	(0.24664898972228338, 0.020627948889953597)
	(0.24903967412853714, 0.021047751455084654)
	(0.24388674542677477, 0.020612089804713504)
	(0.24662056076790442, 0.021095061839489705)
	(0.2474401812369093, 0.02116997428642529)
	(0.2489931883837885, 0.021275467759447136)
	(0.24797718013637207, 0.02055475840798098)
	(0.24575859505884456, 0.021291504995416217)
	(0.2466842972531035, 0.02065731932973761)
	(0.24949858438145045, 0.021881883389321236)
	(0.24721723972010637, 0.021526832925411882)
	(0.24911187849872227, 0.021789484881105246)
	(0.24799673056069987, 0.02147918599349061)
	(0.24593606707688512, 0.020984189841320673)
	(0.247906886999547, 0.02090750923576222)
	(0.24548112610014816, 0.021593525597374916)
	(0.2460933707623029, 0.020904477727385747)
	(0.24718407498809522, 0.02111013887961554)
	(0.24718650277004575, 0.021223427601008853)
	(0.2480653535306135, 0.02201422253275457)
	(0.24431759131556552, 0.021444119572030223)
	(0.2475023261102677, 0.021337922669014128)
	(0.24415476707856532, 0.020789360726013076)
	(0.24754534267029926, 0.021666265316737644)
	(0.24568195276451785, 0.021167828655680582)
	(0.24705715432458317, 0.021332555949761262)
	(0.24757426463514182, 0.02139606796741893)
	(0.24596016187115233, 0.02161172296455077)
	(0.24713794535227615, 0.0208664752206787)
	(0.24740745983379342, 0.021176214311690988)
	(0.2489687194486786, 0.020657180850372894)
	(0.24732654162421067, 0.02121839041984813)
	(0.24930611427928334, 0.020623247953763288)
	(0.24790100791726752, 0.021369752200341242)
	(0.24689880752926446, 0.020527973103903127)
	(0.24938979342683718, 0.020479855942279142)
	(0.24626348587122723, 0.02137453143544341)
	(0.2456117119384697, 0.020488110583813765)
	(0.24845302133151576, 0.02068668815623276)
	(0.24965438369010656, 0.02204433034593943)
	(0.24747174192932453, 0.020258013795128833)
	(0.24631997029222724, 0.021826591600558432)
	(0.24456674924746924, 0.020940629857491445)
	(0.24920317983727575, 0.021899256219252795)
	(0.24773140024664841, 0.021414713463976717)
	(0.2475509628168016, 0.021219163508789468)
	(0.24878104479392207, 0.020588196316356364)
	(0.24687897976337414, 0.021714574071202034)
	(0.24805095408920189, 0.0204055177543098)
	(0.24551485209240895, 0.020972146213477455)
	(0.24410572537862127, 0.020901680457412378)
	(0.24933223642224506, 0.02198904112827692)
	(0.24533005311797554, 0.021451055569912773)
	(0.24693789713601494, 0.020566492261389437)
	(0.2483615992975878, 0.0216791339565177)
	(0.24588668579205192, 0.020900338732967646)
	(0.24578526808278528, 0.021832203780021583)
	(0.24758620371448242, 0.021679141904731365)
	(0.24913954109059747, 0.020746577548302253)
	(0.24633827255195045, 0.02147262198393542)
	(0.2451014275948938, 0.021689855413895988)
	(0.24855881107179673, 0.02140565135077426)
	(0.24705108654303348, 0.021863912706383947)
	(0.24616894307780884, 0.020903970671937225)
	(0.24370760540832934, 0.020220167801603805)
	(0.2463569858675686, 0.021330342792158577)
	(0.24572643828919524, 0.02032495986653133)
	(0.24641773654431928, 0.020770025005754384)
	(0.24809927836351192, 0.021408629609011312)
	(0.24853613534845168, 0.02079529528128331)
	(0.2457559546982764, 0.020578289018098387)
	(0.24502813353184022, 0.021307483407093524)
	(0.24757645971109532, 0.02175670821148886)
	(0.24794877788625833, 0.021948226681809832)
	(0.24656771938863312, 0.02167029474068828)
	(0.24718910080162945, 0.02131108992954199)
	(0.2480395329492369, 0.021014080013002617)
	(0.2450150566005533, 0.021247831787569473)
	(0.24908996141447742, 0.021737287972763454)
	(0.24879090052077296, 0.020547524965956635)
	(0.24725868720787667, 0.02189538919580071)
	(0.24713223876045382, 0.021378242488380984)
	(0.24561531372522397, 0.021649584831691764)
	(0.24656570696743527, 0.021660121287179823)
	(0.24572250622159883, 0.020965473415316724)
	(0.24775307887877115, 0.02065916226210749)
	(0.24690645492951238, 0.021304704191039043)
	(0.2477422062311633, 0.02106548735478357)
	(0.24560969155442167, 0.021056734888978523)
	(0.24688447400783353, 0.02105940753132338)
	(0.24479494924588538, 0.02157567363651153)
	(0.24463129133634182, 0.02165192786669866)
	(0.2465912538494522, 0.021134351744206012)
	(0.24523543165163733, 0.020880949843404877)
	(0.24696066136924505, 0.020562905408596184)
	(0.24798662882049186, 0.020884326131182785)
	(0.24462889259173373, 0.02123007030659468)
	(0.246197534155707, 0.021304495815431847)
	(0.24540170962124191, 0.020635888116319293)
	(0.2495044110919667, 0.02062191808326393)
	(0.249706830418722, 0.021521169617537163)
	(0.24564985354304464, 0.02196600674098133)
	(0.24804738021966444, 0.0217221575148799)
	(0.24745853866506823, 0.02151493464534719)
	(0.24480051660767885, 0.021736716104731275)
	(0.24806047726676267, 0.021095835727348248)
	(0.24905742057636332, 0.02094238865394635)
	(0.24898644905025402, 0.02063197443772286)
	(0.2445865806639097, 0.020865087770713386)
	(0.24959235806941643, 0.021860441063626713)
	(0.24742068135830864, 0.020710408504904194)
	(0.24829179881548685, 0.02038118484078249)
	(0.24769205780600417, 0.02105275536825015)
	(0.246525694769611, 0.021602114597436373)
	(0.24779501741629967, 0.02087439756386208)
	(0.24835054711537158, 0.020756272144642963)
	(0.24956030494464462, 0.02120215682000379)
	(0.2459226527321011, 0.021543991597230222)
	(0.24946847154185092, 0.021017657729699022)
	(0.24527313514155158, 0.020566981693168503)
	(0.24913935158539888, 0.0214048618926665)
	(0.24824697705702428, 0.021008473498993853)
	(0.24448074569388292, 0.020505109151689987)
	(0.24899901751316408, 0.020488129857441837)
	(0.24690371251043225, 0.022018425839763964)
	(0.24542047290469246, 0.021998019696200157)
	(0.245961916023633, 0.021805207518116165)
	(0.24516704525924904, 0.021482664188111237)
	(0.2482561164318407, 0.020660862237884745)
	(0.24594972108646337, 0.020863240745172165)
	(0.2453356349970446, 0.020667075182241246)
	(0.24664871209047967, 0.02065326716877539)
	(0.24944238820368042, 0.020834256821486524)
	(0.24867111192838975, 0.0217896786498829)
	(0.24412896138113302, 0.021375501187606748)
	(0.24869591994000295, 0.021453712006981138)
	(0.2455140662244958, 0.020625504339269157)
	(0.24884352625504463, 0.021052566773917994)
	(0.2457651589094674, 0.020503192699545154)
	(0.24527221639840788, 0.021275647203430177)
	(0.2457601611117106, 0.021100461808692923)
	(0.24868095400838097, 0.02125751463046212)
	(0.2485158440612703, 0.022097038221450076)
	(0.2452525117183344, 0.021742993607755192)
	(0.2453070941164926, 0.02025857850617082)
	(0.2461336499557052, 0.02087280061452393)
	(0.24611511028016453, 0.020624910813081612)
	(0.2444020669783519, 0.020439119708226042)
	(0.24902578277158371, 0.020911766296318596)
	(0.24404263125649883, 0.021675476857331365)
	(0.24568032132851175, 0.021667564008572324)
	(0.24533030587105129, 0.02125695971787856)
	(0.24623753953012534, 0.020702049349675997)
	(0.24747518703877336, 0.02136839463879671)
	(0.24947604688815866, 0.02116525762310315)
	(0.24446406989520705, 0.020825929197062987)
	(0.24461828876037942, 0.020625693355823115)
	(0.24638622227391885, 0.021736569003437452)
	(0.24422856512769472, 0.02090812811248244)
	(0.24940362166611296, 0.021412105840519828)
	(0.24409842757263409, 0.02148412707976969)
	(0.2492199294246883, 0.020965120142743595)
	(0.24650969170877138, 0.021067952360815087)
	(0.24614797352972195, 0.02096970312472511)
	(0.244460273457815, 0.0207492732746369)
	(0.2449435387483223, 0.02091814916980555)
	(0.24787903547579473, 0.020917628282884522)
	(0.24705184881474054, 0.02170233997208856)
	(0.24701273830044007, 0.021454504622028935)
	(0.2484480449956393, 0.02151982390120152)
	(0.24685134193956046, 0.02168243937056586)
	(0.24510038503875686, 0.020321303658235427)
	(0.24577863261754293, 0.020684316664403957)
	(0.24702104116268203, 0.021739264759614536)
	(0.24585954905506113, 0.021221346298227597)
	(0.24411464731955623, 0.02120836426015161)
	(0.24416027706688564, 0.020787740779923063)
	(0.24529918261848324, 0.020995974271343604)
	(0.2439089744790212, 0.021377857020040344)
	(0.24491403956809565, 0.02209397359106564)
	(0.24490301856948266, 0.02040091853543111)
	(0.2467147108558595, 0.020543150612891723)
	(0.24599358456910234, 0.02049617493273668)
	(0.2451112490022439, 0.020576476836110524)
	(0.24588334607072743, 0.021740729560726306)
	(0.24783190218115328, 0.02098220012669211)
	(0.24484146443088695, 0.021439321432545747)
	(0.24653275142878084, 0.021782855089574178)
	(0.24796464499572837, 0.021024332312010736)
	(0.24854154138370618, 0.021939277701575693)
	(0.24450265725671574, 0.02202101785418481)
	(0.2481523328868094, 0.02103140246870602)
	(0.24849745059137412, 0.02150034528040803)
	(0.24439088592402786, 0.0210118057197486)
	(0.24518456684899145, 0.021845296452840276)
	(0.2452769612066059, 0.020723444271159262)
	(0.24651734968015512, 0.020965112295551307)
	(0.24685757642269443, 0.021574441423699527)
	(0.2490151190376876, 0.021105071876961586)
	(0.24809085665695932, 0.021299126154634376)
	(0.2490090392325813, 0.02219530481818617)
	(0.2446221380165673, 0.020282895822996954)
	(0.24690040865874524, 0.020628602035408217)
	(0.24723970440821, 0.02158864195953078)
	(0.24637108048381123, 0.02187940224390419)
	(0.24927432717634013, 0.02176148566348347)
	(0.24764735495904544, 0.021355613126961843)
	(0.2476922038180331, 0.020221543464145814)
	(0.24436169805905597, 0.020583391628371906)
	(0.24753847508550852, 0.020682238228425062)
	(0.2449080065084866, 0.02160815350779578)
	(0.2474676593887578, 0.02104668966499785)
	(0.24795969079713093, 0.021882624324279124)
	(0.2457430941200564, 0.02196258569316277)
	(0.24905211810960243, 0.021914847822414894)
	(0.24535991286352343, 0.02157140163403016)
	(0.2451628430688647, 0.021709244389735524)
	(0.24707149057393554, 0.020483752271435274)
	(0.24687514344503733, 0.020437038067042855)
	(0.24486765367596186, 0.020795207422951833)
	(0.24458116135325048, 0.02137206182592279)
};
\addplot+[draw=none, mark options={fill=green!50!black,draw=green!50!black},only marks,mark size=1,mark=*] coordinates {
	(0.22740436357362054, 0.02194474764705569)
	(0.22698923670466575, 0.021377376680065677)
	(0.22386624867026375, 0.021729315604116537)
	(0.2242063675546246, 0.02134315364248206)
	(0.2245012625764902, 0.02150718422430723)
	(0.2245005133886724, 0.021505780456506578)
	(0.22578672038121272, 0.021679377755038583)
	(0.22687417045291125, 0.021769540827552256)
	(0.22777253852897894, 0.021774411403351433)
	(0.22640465939851045, 0.021801661301405262)
	(0.22492548075730734, 0.02183696552546662)
	(0.22254940753048608, 0.021918071196025957)
	(0.22483138414045056, 0.021483530476166555)
	(0.22680621679830604, 0.021372139019098263)
	(0.2268784969061575, 0.021400287081932236)
	(0.22553001130984446, 0.021814644640660616)
	(0.22429234020013755, 0.021427556843854337)
	(0.22772148577567275, 0.02136170207170719)
	(0.22296156217918628, 0.021584990518998454)
	(0.2225779432655458, 0.02156908018595555)
	(0.22754682179592842, 0.021779134131599227)
	(0.2251010663975531, 0.021763175843543328)
	(0.22620098846537934, 0.021447918988018025)
	(0.22438118318236513, 0.021441721540674866)
	(0.22551189655761136, 0.021566614645681315)
	(0.22787268368088556, 0.021906618088679622)
	(0.2224291815918809, 0.0218560258082683)
	(0.2255095825002285, 0.021356686368473665)
	(0.22523555381201354, 0.02146811998514596)
	(0.22711996404638193, 0.02168331218661531)
	(0.22290004450344356, 0.021347080202599657)
	(0.22265228856009647, 0.021939597579762462)
	(0.22359046847998493, 0.021651113844548712)
	(0.22686614512085646, 0.02189538779366075)
	(0.22330595649211124, 0.021393745706667088)
	(0.22557675935599567, 0.02172840346800196)
	(0.22400203140293418, 0.021514231621954684)
	(0.2267348127930181, 0.021540751245710046)
	(0.2265719833118639, 0.021657286051871345)
	(0.22616825560216997, 0.02183087560028607)
	(0.2229142681215619, 0.021924708656013075)
	(0.22457236171264947, 0.021356684922602467)
	(0.22650498229575192, 0.021471158258165302)
	(0.22643621865565794, 0.021409526825695753)
	(0.22305620300447718, 0.021770556559557397)
	(0.2250446910258247, 0.021405497567525297)
	(0.22348005501073892, 0.021489459060586753)
	(0.22270067329154983, 0.021849979525793508)
	(0.22500883449468284, 0.02152523789357491)
	(0.22581534810369416, 0.02193958671459398)
	(0.22676356862674762, 0.021540952332888115)
	(0.2255792608848916, 0.021901567740543755)
	(0.2260628663657464, 0.021857478324515244)
	(0.22366116403436379, 0.02175226424201871)
	(0.22289341997458376, 0.021908458364506053)
	(0.22425135478377245, 0.021357670847259497)
	(0.2247791335166247, 0.02157907344365073)
	(0.2262618050364008, 0.021394738792558113)
	(0.2258829303715565, 0.02187758981174437)
	(0.22391340604525917, 0.02184362439174891)
	(0.22323118916186066, 0.02165920914217581)
	(0.2248975460817775, 0.021874625073441057)
	(0.2268457954617346, 0.021542076693482504)
	(0.22571984853205543, 0.02184386137553022)
	(0.22607716568413194, 0.021545119078606866)
	(0.22512582160882794, 0.0215564022413412)
	(0.2269736720921696, 0.021693819957018634)
	(0.22385711766072713, 0.021339710677118134)
	(0.22266765868229663, 0.021802353530017162)
	(0.22258548478061116, 0.02144598115091698)
	(0.2256239073414556, 0.02156609732632852)
	(0.22730436384900882, 0.02182932910154166)
	(0.22310377889552818, 0.021780013686208036)
	(0.2253686079859576, 0.021916732276745524)
	(0.2225579020204134, 0.02159220818825597)
	(0.22633699865018156, 0.021362269717617222)
	(0.22519214059326412, 0.02193879088708062)
	(0.2244027316715657, 0.021929150401191734)
	(0.22283041506813928, 0.021381639727383408)
	(0.22405758037497767, 0.021465362241074704)
	(0.226110756835672, 0.021779123806016293)
	(0.22722963943129001, 0.02171698698118413)
	(0.22666134229412857, 0.021628433886318698)
	(0.22747374503406254, 0.02144732680238153)
	(0.22701821567720404, 0.021591716996398877)
	(0.2243495308589484, 0.02135495022422805)
	(0.22398704141530937, 0.021762445690338958)
	(0.22546737738489245, 0.021562405029172428)
	(0.2258341959771157, 0.021637935191381345)
	(0.22457530756647082, 0.021907772085348304)
	(0.22682795802174213, 0.021582121485467003)
	(0.2277602488872006, 0.021737312421760108)
	(0.22505922779029838, 0.02190319816195782)
	(0.2249075025718444, 0.021622436550810864)
	(0.2228970781187054, 0.02191103346831348)
	(0.22614547451296957, 0.02169414393344309)
	(0.2243064894779953, 0.021616945542581794)
	(0.22663907582545206, 0.02182844046284698)
	(0.22555808590120108, 0.02193317085701229)
	(0.2261000785329396, 0.021459423001717468)
	(0.22329747989443138, 0.021599510791352288)
	(0.22329969792453214, 0.021869427346764097)
	(0.22499602151275774, 0.0216687052579243)
	(0.2231527543065609, 0.021689533500978003)
	(0.22433162516438634, 0.02183813694159444)
	(0.22753695017865685, 0.0215474286541053)
	(0.22507443272159647, 0.02149601631071666)
	(0.225597973373535, 0.021665492573071515)
	(0.22455486916299786, 0.021943198953540237)
	(0.22251371409736267, 0.021421491564815858)
	(0.22676878171766326, 0.02155347848990113)
	(0.22502510954373348, 0.021431371064956677)
	(0.22555946257699788, 0.02193226989278413)
	(0.22586731965859172, 0.0214756504142961)
	(0.22395817333836557, 0.02135025326610707)
	(0.2227674022398648, 0.021720562721830535)
	(0.22542537740923288, 0.02182634464596008)
	(0.22535478629673136, 0.02183558149155143)
	(0.22457326203630354, 0.021356710512309536)
	(0.22297291558218546, 0.021630942818871637)
	(0.22362347468432092, 0.021533326443428837)
	(0.22530007035612204, 0.021874690535404164)
	(0.22589692373196213, 0.02137000208211967)
	(0.22247626293476455, 0.02179042277320583)
	(0.2278076904799018, 0.02169690801475817)
	(0.2276516217451842, 0.021420933686643218)
	(0.22655479596223282, 0.021479923499376337)
	(0.22248272915273976, 0.02159206396496907)
	(0.22407065068412482, 0.021669043127080144)
	(0.22607916945347528, 0.02177549576645982)
	(0.22356215131609902, 0.02180558858735225)
	(0.22433871820204243, 0.021633067116581084)
	(0.22260535456518063, 0.021522096967312485)
	(0.2256044326743397, 0.021747188486021418)
	(0.22615019325972338, 0.021884350733414184)
	(0.22410512039842165, 0.02189956389243501)
	(0.2245142554758597, 0.021930916968325966)
	(0.22437804223440863, 0.021415264249903025)
	(0.2231156073247636, 0.021883499345452938)
	(0.22628009332802732, 0.021686140387530545)
	(0.22461780407656654, 0.021464825146095583)
	(0.22503792640241585, 0.02141246543624993)
	(0.22474951057549525, 0.021768931209931026)
	(0.22605410015216582, 0.021559262312237273)
	(0.2259052758811715, 0.02140896676843162)
	(0.2270801647628941, 0.021448122210539124)
	(0.2242973099018873, 0.021651844040881173)
	(0.2263639244197731, 0.02155594816330214)
	(0.22379921485529797, 0.021882249293701056)
	(0.22571501346482256, 0.021792671489627068)
	(0.2264526183179485, 0.02172595830964546)
	(0.22714529140342604, 0.021790044994107197)
	(0.22577250985802025, 0.021387135211910234)
	(0.2261358945700503, 0.02176287988835514)
	(0.22762547928007665, 0.021588546595979717)
	(0.22403070268546071, 0.021786658013509676)
	(0.22594173231784334, 0.02144998097143299)
	(0.22483324233605925, 0.021391981039830598)
	(0.225682565159123, 0.02186477361011451)
	(0.22400500873043208, 0.021515525522504783)
	(0.22659231100129149, 0.021735535654703285)
	(0.22589840249638882, 0.021920983212493743)
	(0.22780569497324127, 0.021622743051462685)
	(0.2258048674827187, 0.02186923431687595)
	(0.22548454127258208, 0.021475259168446618)
	(0.2251909720164567, 0.021835838568176686)
	(0.22787730108403506, 0.02159794064490108)
	(0.22240290556183812, 0.021933612212095752)
	(0.2277688516469271, 0.021556892355983712)
	(0.22674088140885448, 0.021512681919201963)
	(0.22576693180314356, 0.021759193401544395)
	(0.2248624939330193, 0.021889602673214797)
	(0.22331756511986936, 0.021857000777739555)
	(0.22567793107260498, 0.0217511563622331)
	(0.22407867582417754, 0.021872146863536238)
	(0.22503434989366894, 0.021778345554943256)
	(0.22598159546828972, 0.021830343245630646)
	(0.22353651266509295, 0.021685098005069427)
	(0.22242683965227358, 0.02183431814118527)
	(0.22654087495062714, 0.021660632442824405)
	(0.22714849249511349, 0.02184320476555882)
	(0.2276774213694855, 0.021578900351844888)
	(0.22401762348955565, 0.02150283426444035)
	(0.22443976362330054, 0.021833258340770784)
	(0.2243315845071037, 0.021771152144608574)
	(0.2275172101127092, 0.021565793441302356)
	(0.22590566452735167, 0.02145593034346153)
	(0.22598939236003715, 0.02190960137023286)
	(0.22244163142913748, 0.021928223462649953)
	(0.22478173497049475, 0.02166761089813573)
	(0.22764558131732857, 0.021961815788124218)
	(0.22642826912586203, 0.02184091678709397)
	(0.2235648528709931, 0.02174408734221847)
	(0.22428838756419658, 0.021781100630591275)
	(0.2240830181686031, 0.021701946778197888)
	(0.2262344605396298, 0.021646411255255377)
	(0.22337590737163132, 0.021909440589204958)
	(0.2250469131179127, 0.021727896079096615)
	(0.22520499108094696, 0.02147561069592258)
	(0.2274875235489285, 0.021882992564174822)
	(0.2269451891280649, 0.021530005533363452)
	(0.22340799147220403, 0.021478036121665354)
	(0.2271806737801899, 0.021457282964336532)
	(0.22639139390008214, 0.02171317372611972)
	(0.22716349978204517, 0.021486878454827423)
	(0.2262943792806697, 0.021844105171597593)
	(0.22715650888601138, 0.021551708478738704)
	(0.22359743092555356, 0.021530727446775273)
	(0.22429112210826121, 0.0216327549747495)
	(0.22264499037920765, 0.021726865461462217)
	(0.2231242853003817, 0.02188498213053444)
	(0.2249197708869233, 0.02188038097781519)
	(0.22648194616591738, 0.021365790235863134)
	(0.22343961683884678, 0.02173747467777732)
	(0.22713196740055486, 0.021760284128979523)
	(0.2257212026107149, 0.02192861022438316)
	(0.22432225333844993, 0.021544117505199033)
	(0.2269408303689818, 0.02145646607185765)
	(0.22435352112696777, 0.02151855180224071)
	(0.22566309499496995, 0.0214257600356973)
	(0.22551430371740047, 0.021854290274388028)
	(0.22490087974108122, 0.021539425100930924)
	(0.22678186453693205, 0.021542014250122324)
	(0.22778286184795243, 0.02174046775989341)
	(0.2277666754277549, 0.021789614579993333)
	(0.22278769961052483, 0.021614939394480507)
	(0.22653297798860375, 0.02186950657354593)
	(0.22300334942018107, 0.02178369830189472)
	(0.2224185433148724, 0.02161748518661406)
	(0.2228355520341292, 0.021495476812297118)
	(0.22737733834564386, 0.021571376950335154)
	(0.2248995934397959, 0.021761445093480238)
	(0.22274991758338158, 0.021828851852025412)
	(0.2267103046430832, 0.021725873499878985)
	(0.22236141124340678, 0.02178579433810668)
	(0.2238123283882674, 0.021510453400346085)
	(0.22444590189289307, 0.0216018477303367)
	(0.22626678251175839, 0.021588076575596204)
	(0.22526904744162154, 0.02174298656252506)
	(0.22676451004336162, 0.02154280603680603)
	(0.22547203388523032, 0.021484338533414214)
	(0.22660288357165853, 0.021668362185408536)
	(0.22555077615487282, 0.021853620040489373)
	(0.22476308071345585, 0.021595074080385426)
	(0.22759755284632732, 0.02157618226210473)
	(0.2254680708688453, 0.02174655817086882)
	(0.2259681535782555, 0.021399205352368784)
	(0.22764857690437332, 0.02151203097354137)
	(0.22399611329491817, 0.021865762263703684)
	(0.22326001806882087, 0.021577963942381085)
	(0.22272291274615835, 0.02168465538525231)
	(0.22381852035852912, 0.021600464802625917)
	(0.22278282551022419, 0.021654571213997116)
	(0.227652601236297, 0.021353809889395423)
	(0.22284687440582526, 0.021354222000681843)
	(0.2230638522072315, 0.021939118737732112)
	(0.22439792439962125, 0.021651757125071332)
	(0.22260363261211666, 0.021878115748474574)
	(0.2236882353618815, 0.021940631282630317)
	(0.2254924797037514, 0.02136678194623288)
	(0.22404906226955876, 0.021359914746495515)
	(0.2234240754286001, 0.02194358355729488)
	(0.22551569404895586, 0.02169184887606065)
	(0.22640404689027363, 0.021397522591795486)
	(0.22431690315043093, 0.02183016699633527)
	(0.22608853912687132, 0.021841992364528485)
	(0.22555727494125435, 0.021573179530207723)
	(0.2245542251784087, 0.02150344219772949)
	(0.22302755784468126, 0.021585653737313373)
	(0.22523498813050027, 0.02148598841847088)
	(0.22425585892134398, 0.021775913093795177)
	(0.22539635851440068, 0.021570697189715965)
	(0.22548391599903703, 0.021420565955371205)
	(0.22566736020793726, 0.02158060179476668)
	(0.22672031269924275, 0.021850592929901887)
	(0.22321656322888925, 0.02153473861775371)
	(0.22540631749227838, 0.02145720202301941)
	(0.22731109219283657, 0.021647665443960174)
	(0.22240472150539398, 0.02195100869735481)
	(0.22519481114289278, 0.021454178579390595)
	(0.2274403075767537, 0.02159936655178343)
	(0.22246660732123072, 0.021420138105544054)
	(0.2251136614108639, 0.02150689935704053)
	(0.2256264715531769, 0.02181370968373242)
	(0.22725983797703186, 0.02173335040675664)
	(0.22627644746429884, 0.02170073267207322)
	(0.22417057502000579, 0.02158802003883878)
	(0.22520063440867227, 0.02148366284443123)
	(0.22755345950677677, 0.02194512487467369)
	(0.22527800885987637, 0.021939230860230006)
	(0.2273217121506132, 0.02179016634810907)
	(0.22620796430378828, 0.021788766256911576)
	(0.2242630161611433, 0.021673050915741815)
	(0.22610980877127523, 0.021797078228271767)
	(0.22353652344566832, 0.02194460265447984)
	(0.22424686726555415, 0.021846503496748743)
	(0.2256725261691148, 0.021511548818980434)
	(0.2258045929080063, 0.021381909862039444)
	(0.2261513171800543, 0.02191403635055922)
	(0.22274982938822757, 0.02156776192733794)
	(0.22563789244792, 0.021864433662347693)
	(0.2226912143338895, 0.02146355274467581)
	(0.22569230595068585, 0.021853036719613417)
	(0.2240920434654303, 0.021589909299087554)
	(0.22544660975983227, 0.021610544564750887)
	(0.22563485976332892, 0.0219394048718129)
	(0.22424486633846452, 0.021715295532687806)
	(0.2254192162691896, 0.02171872908308656)
	(0.22569496705610034, 0.021712492777693066)
	(0.22728606858025427, 0.021682650868424337)
	(0.22553152045094643, 0.021795021173264224)
	(0.22749794417326166, 0.02180817010602167)
	(0.22614624841075848, 0.02175475950650905)
	(0.22547714462715418, 0.02142166290211029)
	(0.22781470507955318, 0.021575088347284006)
	(0.2249211630216836, 0.02134232284954362)
	(0.22420728329250203, 0.021404428645967657)
	(0.22668313179542435, 0.021769889536091422)
	(0.22775475187103167, 0.021899631819074887)
	(0.2260633168606901, 0.02140842506863444)
	(0.22436833172418136, 0.02195163856804589)
	(0.2227873477671578, 0.021779401480311426)
	(0.22735224933582934, 0.02185093050144639)
	(0.22633219911923705, 0.021399201127411352)
	(0.225929881020092, 0.02162108179670959)
	(0.22700465021406704, 0.021776394579855033)
	(0.22502153610315298, 0.021857443660221146)
	(0.22671225333617073, 0.021338700753031153)
	(0.2241573850419853, 0.021357467050423663)
	(0.22255531481637988, 0.021550410562241384)
	(0.2276176082555923, 0.021714628166652768)
	(0.22390076372206563, 0.021429289395909913)
	(0.22554891778849304, 0.021388979347521916)
	(0.22666707199488767, 0.021694527302700134)
	(0.22401920023132968, 0.02186748556072223)
	(0.22400694708850338, 0.02177832099428189)
	(0.22573484706751468, 0.02185135664696774)
	(0.2277300320814143, 0.021409509009183165)
	(0.22472588465734736, 0.021612387894603104)
	(0.22366961652806627, 0.02143181106682755)
	(0.22715717663660226, 0.021401634435194453)
	(0.2253426682192891, 0.02170841832374438)
	(0.22461392941229005, 0.021555013665518805)
	(0.22235964039826744, 0.021347965010061903)
	(0.2245418141012255, 0.0218151717663431)
	(0.22432149255947798, 0.021404945729717267)
	(0.2247386214679136, 0.02167911507640569)
	(0.22669469095622877, 0.02140458740728314)
	(0.22717538374769586, 0.021360751600755814)
	(0.2242622361578099, 0.021493718540466507)
	(0.22313446753248137, 0.021893665999358852)
	(0.2258795104838023, 0.021696949227293025)
	(0.2262214948988036, 0.02172728298745471)
	(0.22482697932744936, 0.02174074006118375)
	(0.22538508269569865, 0.021804018105930793)
	(0.2264628713813636, 0.021576661567873308)
	(0.22306575452563204, 0.021949302074921268)
	(0.22735880964852248, 0.021731151765954942)
	(0.22736872703610644, 0.02142217348466652)
	(0.2256420700334199, 0.021616617174456756)
	(0.2252041475284031, 0.021928091232050723)
	(0.22378538792633468, 0.021829925798889278)
	(0.22480374719075966, 0.02176195977667561)
	(0.22437268464773344, 0.0213498215738654)
	(0.22639450534149802, 0.02135857353727313)
	(0.22544631449548272, 0.02146014043402965)
	(0.22634723728754108, 0.021394968943622215)
	(0.2242143480127223, 0.021395343541699367)
	(0.22546033723698783, 0.021424136770845703)
	(0.2232846820174556, 0.021510267228429785)
	(0.22268595530768817, 0.02194533602865366)
	(0.22481793486124363, 0.021773318988208584)
	(0.22358694163515053, 0.021648490016486815)
	(0.22520916412293832, 0.021751497246306747)
	(0.22630947093524634, 0.021677157885245518)
	(0.2227905474689633, 0.02183834512277043)
	(0.22457638793891482, 0.021621146216792172)
	(0.22387980749388006, 0.021521902127361853)
	(0.22770827496833956, 0.021796136123627134)
	(0.22786411913466936, 0.021842711284052633)
	(0.2239313715239989, 0.021718482019045755)
	(0.22630800072554325, 0.021739379494121202)
	(0.22602942836016257, 0.02142911030490565)
	(0.22294124969129536, 0.021859266916383475)
	(0.22654787131365847, 0.02151260595310421)
	(0.2276790825590601, 0.021378338017303225)
	(0.22748835332916972, 0.021498095721084305)
	(0.22305302336387575, 0.021533557300033954)
	(0.22783531817978922, 0.021757039889627214)
	(0.22584782747819976, 0.02157285388010889)
	(0.22672242270936677, 0.02156937610612008)
	(0.22611777265503805, 0.021574285150966137)
	(0.22483853655620017, 0.021687158213410836)
	(0.22609201603378454, 0.02170300138251512)
	(0.22647083649517122, 0.02187971062020037)
	(0.2276388395040715, 0.021921465440573135)
	(0.22404333814494312, 0.021879314587157972)
	(0.2278481963362391, 0.021620275205611824)
	(0.22350531314697056, 0.021767821994581008)
	(0.2275869937384751, 0.021552357846923786)
	(0.2265836011762505, 0.02166337588077378)
	(0.22287368262374682, 0.021607063070136108)
	(0.22733527444219098, 0.02166374307097311)
	(0.22503087646483066, 0.02187283604560158)
	(0.223648502757772, 0.021771970146920446)
	(0.22423101938090656, 0.02173089664272642)
	(0.22338328986630165, 0.021783755392947395)
	(0.22654454274924543, 0.021711573682595266)
	(0.22406015822808445, 0.02188956285837893)
	(0.22394447794803768, 0.021391157049006924)
	(0.22497496772942607, 0.021673744361053612)
	(0.2278323463932622, 0.021610041810418222)
	(0.22698537039603242, 0.021685741532357323)
	(0.2224995442810197, 0.02162941710011333)
	(0.22719406659733674, 0.021501853342666206)
	(0.22401848791329326, 0.02149557831120256)
	(0.22732374000004357, 0.021519786255001054)
	(0.2242710426961104, 0.021494116213357)
	(0.22340874910509897, 0.021863467293308913)
	(0.22439337378618907, 0.02136678732552152)
	(0.2270349087153099, 0.021646045293071063)
	(0.22670264924899422, 0.021813194812276088)
	(0.22368222761787457, 0.021570284100459823)
	(0.22388370966975782, 0.02142338444673478)
	(0.22428079647186253, 0.021852853483842655)
	(0.224517900379827, 0.021597209900337543)
	(0.22286646720111505, 0.021535599777236836)
	(0.2271610940918874, 0.02186468867969632)
	(0.22242089853387279, 0.021621732722626043)
	(0.22379094437586952, 0.021889376952642217)
	(0.22398615179874315, 0.021344154072308136)
	(0.2243681144844276, 0.021869425045697735)
	(0.2257843762274476, 0.02169081081132577)
	(0.22785974532095826, 0.02161630156720041)
	(0.22297658538736806, 0.021487484507838998)
	(0.22285181718151256, 0.02176647157886686)
	(0.2245462405052656, 0.021839981768653254)
	(0.22236352915372315, 0.021865035973971582)
	(0.22779062901870373, 0.021612992647409242)
	(0.22253521515340657, 0.021563212419227505)
	(0.22748180450432293, 0.02173812492036536)
	(0.2247337831056406, 0.021775908603130775)
	(0.2242040133994271, 0.02194396013029485)
	(0.22303558383532218, 0.021424689622492835)
	(0.2233727910052529, 0.021570747743069416)
	(0.22649366813568683, 0.0213853673401079)
	(0.22560841427288778, 0.021443434541852758)
	(0.22532246792711116, 0.02169027037332892)
	(0.2267400213661486, 0.021708023629490713)
	(0.22534820735895145, 0.021503134580609016)
	(0.223740989908894, 0.02135939512986287)
	(0.22405469739112258, 0.021723935226420354)
	(0.22555643468884445, 0.02146460647383759)
	(0.2243446939988574, 0.021514855056203725)
	(0.22270671315842816, 0.02140793416112808)
	(0.22242882976635786, 0.02173144730052778)
	(0.2234696429687819, 0.02182953964970134)
	(0.22241543881135306, 0.02149353566766816)
	(0.22301806013713163, 0.02189597943096401)
	(0.2234704706648048, 0.021432547904677853)
	(0.22515774323879742, 0.02155696761706209)
	(0.2245844591713779, 0.02140912539772444)
	(0.2234070892897925, 0.021704159712451408)
	(0.22398171455133997, 0.02190163151938745)
	(0.22610251682858715, 0.02172938535256613)
	(0.2233106980240413, 0.02153076640684566)
	(0.22464692824653906, 0.021885823182241786)
	(0.22644832456414563, 0.021516320431582737)
	(0.22661065168333414, 0.02193088970037204)
	(0.22257301140296348, 0.021929645853752257)
	(0.22624722181555854, 0.021905111071250878)
	(0.22682652261779296, 0.021670927973581174)
	(0.22304185882912064, 0.021349027094907224)
	(0.22347242424748007, 0.021712142601511393)
	(0.22361344471113181, 0.02166351649547408)
	(0.2250999439853935, 0.021417405694761603)
	(0.22504076991968244, 0.021816806503011998)
	(0.2272376419582676, 0.021777477079420023)
	(0.22621255208266713, 0.021878304574292178)
	(0.2271045876912499, 0.021904451541331423)
	(0.22320043487348726, 0.021421703143080045)
	(0.22528747246295588, 0.02161293619578935)
	(0.22557700693996882, 0.021662697468241184)
	(0.22466613887005868, 0.021704941613752547)
	(0.22757030051322372, 0.02170402666311642)
	(0.226067993955819, 0.021579361003226424)
	(0.22631555242787962, 0.0213766513901535)
	(0.22269482084838396, 0.021666877210672005)
	(0.2258735032907526, 0.02166497179475594)
	(0.22309817372153398, 0.021809832786952605)
	(0.22608850026035776, 0.021379159128400046)
	(0.22606318030575911, 0.021896510491371816)
	(0.2239835896376194, 0.021759504482436978)
	(0.22728280501405987, 0.021769313095542572)
	(0.2238344552941666, 0.02152545756935684)
	(0.22347125012715816, 0.02169159294170654)
	(0.22555307085092619, 0.021518419723009347)
	(0.2254978857762831, 0.021377257668754222)
	(0.22313305044275936, 0.021734603233202518)
	(0.22305993237762134, 0.021521228975629143)
};
\addplot+ [mark = {diamond*}, mark size = {3}, cyan,thick,solid,mark options={fill=cyan}]coordinates {
	(0.23899074074074075, 0.020714772727272728)
	(0.24453703703703702, 0.020714772727272728)
	(0.24453703703703702, 0.021338636363636365)
	(0.23899074074074075, 0.021338636363636365)
	(0.23899074074074075, 0.020714772727272728)
};
\addplot+ [mark = {diamond*}, mark size = {3}, cyan,thick,solid,mark options={fill=cyan}]coordinates {
	(0.24453703703703702, 0.02009090909090909)
	(0.2500833333333333, 0.02009090909090909)
	(0.2500833333333333, 0.020714772727272728)
	(0.24453703703703702, 0.020714772727272728)
	(0.24453703703703702, 0.02009090909090909)
};
\addplot+ [mark = {diamond*}, mark size = {3}, cyan,thick,solid,mark options={fill=cyan}]coordinates {
	(0.24453703703703702, 0.020714772727272728)
	(0.2500833333333333, 0.020714772727272728)
	(0.2500833333333333, 0.021338636363636365)
	(0.24453703703703702, 0.021338636363636365)
	(0.24453703703703702, 0.020714772727272728)
};
\addplot+ [mark = {diamond*}, mark size = {3}, cyan,thick,solid,mark options={fill=cyan}]coordinates {
	(0.23899074074074075, 0.021338636363636365)
	(0.24453703703703702, 0.021338636363636365)
	(0.24453703703703702, 0.0219625)
	(0.23899074074074075, 0.0219625)
	(0.23899074074074075, 0.021338636363636365)
};
\addplot+ [mark = {diamond*}, mark size = {3}, cyan,thick,solid,mark options={fill=cyan}]coordinates {
	(0.24453703703703702, 0.021338636363636365)
	(0.2500833333333333, 0.021338636363636365)
	(0.2500833333333333, 0.0219625)
	(0.24453703703703702, 0.0219625)
	(0.24453703703703702, 0.021338636363636365)
};
\addplot+ [mark = {diamond*}, mark size = {3}, cyan,thick,solid,mark options={fill=cyan}]coordinates {
	(0.23899074074074075, 0.0219625)
	(0.24453703703703702, 0.0219625)
	(0.24453703703703702, 0.022586363636363637)
	(0.23899074074074075, 0.022586363636363637)
	(0.23899074074074075, 0.0219625)
};
\addplot+ [mark = {diamond*}, mark size = {3}, cyan,thick,solid,mark options={fill=cyan}]coordinates {
	(0.24453703703703702, 0.0219625)
	(0.2500833333333333, 0.0219625)
	(0.2500833333333333, 0.022586363636363637)
	(0.24453703703703702, 0.022586363636363637)
	(0.24453703703703702, 0.0219625)
};
\end{axis}

\end{tikzpicture}

%% file: MC_reach.tex
\begin{tikzpicture}[]
\begin{axis}[height={7.0cm}, width={8.5cm},ylabel = {Maximum Steps to Reach Goal}, xlabel = {$w$}]\addplot+ [mark = {*}, only marks = {true}]coordinates {
(0.0, 127.0)
(0.05, 149.0)
(0.1, 165.0)
(0.15, 216.0)
(0.2, 237.0)
(0.25, 352.0)
};
\addlegendentry{Reachability}
\addplot+ [mark = {square*}, only marks = {true}]coordinates {
(0.0, 100.0)
(0.05, 102.0)
(0.1, 103.0)
(0.15, 106.0)
(0.2, 111.0)
(0.25, 117.0)
(0.3, 125.0)
(0.35, 133.0)
(0.4, 191.0)
(0.45, 207.0)
(0.5, 296.0)
};
\addlegendentry{MC worst}
\addplot+ [mark = {diamond*}, only marks = {true}, mark size = {3}]coordinates {
(0.0, 100.0)
(0.05, 99.0)
(0.1, 100.0)
(0.15, 100.0)
(0.2, 100.0)
(0.25, 100.0)
(0.3, 100.0)
(0.35, 101.0)
(0.4, 101.0)
(0.45, 101.0)
(0.5, 101.0)
(0.6, 102.0)
(0.8, 106.0)
(1.0, 105.0)
};
\addlegendentry{MC random}
\end{axis}

\end{tikzpicture}

%% file: VerticalCAS_ReachableRow.tex
\begin{tikzpicture}[]
\begin{groupplot}[height=4.4cm, width=4.4cm, group style={horizontal sep=0.35cm, group size=5 by 1}]
\nextgroupplot [ylabel = {$h$ (ft)}, title = {$\tau=$40 s}, xlabel = {$\dot{h}_0$ (ft/min)}, , enlargelimits = false, axis on top]\addplot [point meta min=0, point meta max=1] graphics [xmin=-2500, xmax=2500, ymin=-1000, ymax=1000] {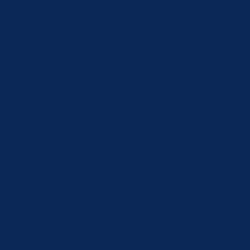};
\addplot+ [red,dashed,no marks]coordinates {
(-2500, -100)
(2500, -100)
};
\addplot+ [red,dashed,no marks]coordinates {
(-2500, 100)
(2500, 100)
};
\nextgroupplot [ylabel = {}, title = {$\tau=$18 s}, xlabel = {$\dot{h}_0$ (ft/min)}, yticklabels={,,}, scaled y ticks=false, enlargelimits = false, axis on top]\addplot [point meta min=0, point meta max=1] graphics [xmin=-2500, xmax=2500, ymin=-1000, ymax=1000] {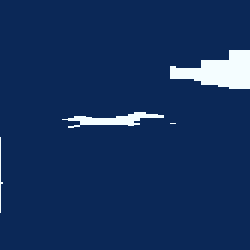};
\addplot+ [red,dashed,no marks]coordinates {
(-2500, -100)
(2500, -100)
};
\addplot+ [red,dashed,no marks]coordinates {
(-2500, 100)
(2500, 100)
};
\nextgroupplot [ylabel = {}, title = {$\tau=$12 s}, xlabel = {$\dot{h}_0$ (ft/min)}, yticklabels={,,}, scaled y ticks=false, enlargelimits = false, axis on top]\addplot [point meta min=0, point meta max=1] graphics [xmin=-2500, xmax=2500, ymin=-1000, ymax=1000] {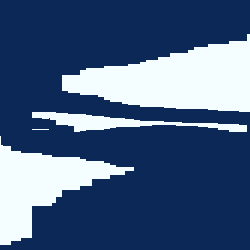};
\addplot+ [red,dashed,no marks]coordinates {
(-2500, -100)
(2500, -100)
};
\addplot+ [red,dashed,no marks]coordinates {
(-2500, 100)
(2500, 100)
};
\nextgroupplot [ylabel = {}, title = {$\tau=$6 s}, xlabel = {$\dot{h}_0$ (ft/min)}, yticklabels={,,}, scaled y ticks=false, enlargelimits = false, axis on top]\addplot [point meta min=0, point meta max=1] graphics [xmin=-2500, xmax=2500, ymin=-1000, ymax=1000] {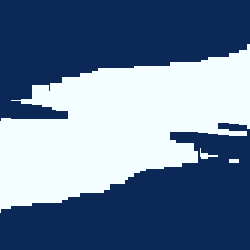};
\addplot+ [red,dashed,no marks]coordinates {
(-2500, -100)
(2500, -100)
};
\addplot+ [red,dashed,no marks]coordinates {
(-2500, 100)
(2500, 100)
};
\nextgroupplot [ylabel = {}, title = {$\tau=$0 s}, xlabel = {$\dot{h}_0$ (ft/min)}, yticklabels={,,}, scaled y ticks=false, enlargelimits = false, axis on top]\addplot [point meta min=0, point meta max=1] graphics [xmin=-2500, xmax=2500, ymin=-1000, ymax=1000] {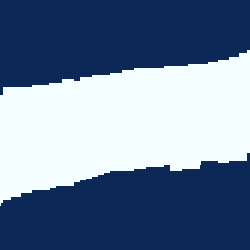};
\addplot+ [red,dashed,no marks]coordinates {
(-2500, -100)
(2500, -100)
};
\addplot+ [red,dashed,no marks]coordinates {
(-2500, 100)
(2500, 100)
};

\node[text=white] at (0,750)  {\footnotesize Reachable};
\node[text=black] at (900,280)  {\footnotesize Unreachable};
\node[text=red] at (-700,-240)  {\footnotesize Unsafe Region};

\end{groupplot}

\end{tikzpicture}

%% file: VertCAS_Policy.tex
\begin{tikzpicture}[]
\begin{groupplot}[group style={horizontal sep=4cm, vertical sep=2.5cm, group size=2 by 2}]
\nextgroupplot [height = {6.9cm}, ylabel = {$h$ (ft)}, xmin = {0.0}, xmax = {40.0}, ymax = {1000.0}, xlabel = {$\tau$ (s)}, ymin = {-1000.0}, width = {8.2cm}, enlargelimits = false, axis on top]\addplot [point meta min=1, point meta max=9] graphics [xmin=0.0, xmax=40.0, ymin=-1000.0, ymax=1000.0] {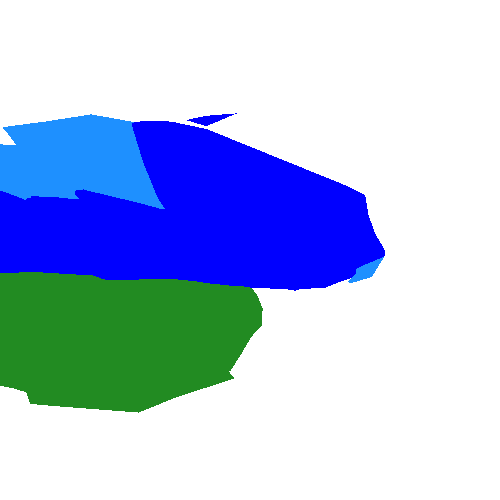};

\node[text=black] at (30,-500)  {\footnotesize COC};
\node[text=white] at (10,-300)  {\footnotesize SCL1500};
\node[text=white] at (20,170)  {\footnotesize SDES1500};
\node[text=white] at (6,350)  {\footnotesize DES1500};

\end{groupplot}

\end{tikzpicture}

%% file: VerticalCAS_ReachablePlot.tex
\begin{tikzpicture}[]

\begin{groupplot}[group style={horizontal sep=0.5cm, group size=2 by 1, vertical sep=1.9cm}, width = 4.5cm, height = 4.5cm]
\nextgroupplot[ylabel = {$h$ (ft)}, xlabel = {$\dot{h}_0$ (ft/min)},enlargelimits = false, axis on top]\addplot [point meta min=0, point meta max=1] graphics [xmin=-2500, xmax=2500, ymin=-1000, ymax=1000] {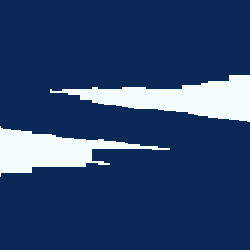};
\addplot+ [red,dashed,thick,no marks]coordinates {
(-2500, -100)
(2500, -100)
};
\addplot+ [red,dashed,thick,no marks]coordinates {
(-2500, 100)
(2500, 100)
};

\nextgroupplot[xlabel = {$\dot{h}_0$ (ft/min)}, enlargelimits = false, axis on top,yticklabels={,,}, scaled y ticks=false,]\addplot [point meta min=0, point meta max=1] graphics [xmin=-2500, xmax=2500, ymin=-1000, ymax=1000] {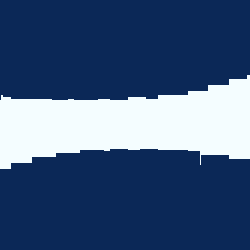};
\addplot+ [red,thick,dashed,no marks]coordinates {
	(-2500, -100)
	(2500, -100)
};
\addplot+ [red,thick,dashed,no marks]coordinates {
	(-2500, 100)
	(2500, 100)
};
\end{groupplot}

\end{tikzpicture}